\documentclass[lettersize,journal]{IEEEtran}
\usepackage{amsmath,amsfonts}
\usepackage{algorithmic}
\usepackage{algorithm}
\usepackage{array}
\usepackage[caption=false,font=normalsize]{subfig}

\usepackage{textcomp}
\usepackage{stfloats}
\usepackage{wrapfig}
\usepackage{float}
\usepackage{url}
\usepackage{verbatim}
\usepackage{graphicx}
\usepackage{booktabs}
\usepackage{cite}
\usepackage{afterpage}

\usepackage{tikz,xcolor}
\usepackage[implicit=false]{hyperref}
\usepackage[english]{babel}
\hypersetup{hidelinks,
	colorlinks=true,
	allcolors=black,
	pdfstartview=Fit,
	breaklinks=true}
	
\definecolor{lime}{HTML}{A6CE39}
\DeclareRobustCommand{\orcidicon}{
\begin{tikzpicture}
\draw[lime, fill=lime] (0,0)
circle[radius=0.16]
node[white]{{\fontfamily{qag}\selectfont \tiny \.{I}D}};
\end{tikzpicture}
\hspace{-2mm}
}
\foreach \x in {A, ..., Z}{%
\expandafter\xdef\csname orcid\x\endcsname{\noexpand\href{https://orcid.org/\csname orcidauthor\x\endcsname}{\noexpand\orcidicon}}
}

\usepackage[capitalise]{cleveref}
\crefname{equation}{Eq.}{Eqs.}          
\crefname{figure}{Fig.}{Figs.}          
\crefname{subfigure}{Fig.}{Figs.}      
\crefname{table}{Table}{Tables}         
\crefname{section}{Section}{Sections}   
\crefname{algorithm}{Algorithm}{Algorithms} 

\begin{document}

\title{High-Precision Modal Analysis of Multimode Waveguides from Amplitudes via Large-Step Nonconvex Optimization\vspace{-1ex}}

\author{Jingtong~Li\hspace{-1.5mm}\orcidA,~Dongting Huang,~Minhui~Xiong~and~Mingzhi~Li\hspace{-1.5mm}\orcidB
\thanks{This work was supapertureed by the NSFC Project under Grants 62231001, 60401012 and 61271052. \textit{(Corresponding author: Mingzhi Li.)}}
\thanks{Jingtong Li, Dongting Huang, and Minhui Xiong are with the School of Electronics, Peking University, Beijing 100871, China (e‑mail: jt\_li@stu.pku.edu.cn).}
\thanks{Mingzhi Li is with the State Key Laboratory of Photonics and Communications and the School of Electronics, Peking University, Beijing 100871, China (e-mail: mileslee@pku.edu.cn).}
}

\markboth{Journal of \LaTeX\ Class Files,~Vol.~14, No.~8, August~2021}%
{Shell \MakeLowercase{\textit{et al.}}: A Sample Article Using IEEEtran.cls for IEEE Journals}

\IEEEpubid{0000--0000/00\$00.00~\copyright~2021 IEEE}

\maketitle

\begin{abstract}

Optimizing multimodal waveguide performance depends on modal analysis; however, existing methods focus predominantly on modal power distribution (MPD) and, limited by experimental hardware and conditions, exhibit low accuracy, poor adaptability, and high computational cost.
This work presents a novel framework for comprehensive modal analysis  (recovering both power and relative phase) using aperture field (AF) and far field (FF) amplitude measurements.  We formulate the modal analysis as a nonconvex optimization problem under a power-normalization constraint and, inspired by recent advances in deep learning, introduce a large-step strategy  to solve it. 
Our method retrieves both the MPD and the modal relative-phase distribution(MRPD). The effectiveness of the proposed method is validated through visualization of the nonconvex optimization process via its loss landscape. Under noiseless conditions, analysis results of 
$93$ electromagnetic modes indicate that the relative amplitude accuracy $\mathrm{MRE_{Modulus}}$, and the phase accuracy $\mathrm{MAE_{Phase}}$, both reach the level of machine precision. 
Through noise simulations of the AF and environmental background, the operational principles of the method are demonstrated under signal-to-noise ratio (SNR) conditions ranging from $10~\mathrm{dB}$ to  $60~\mathrm{dB}$. Experiments further confirm that error suppression is effectively achieved by increasing the number of sampling points, thereby maintaining high accuracy and strong robustness. Within a unified evaluation framework, the absolute amplitude error $\mathrm{MAE_{Modulus}}$, and the phase error $\mathrm{MAE_{Phase}}$, are as low as $1.633\times10^{-8}$ and $0$, respectively. The accuracy is significantly superior to existing methods, while also exhibiting higher computational efficiency.

\end{abstract}

\begin{IEEEkeywords}
Modal analysis, multimode waveguide, nonconvex optimization, phase retrieval, power normalization, gradient descent, large step size, automatic differentiation, deep learning
\end{IEEEkeywords}

\section{Introduction}\label{section:introduction}

\IEEEPARstart{M}{ultimode} waveguides play a pivotal role in high-capacity communications and high-power transmission, and their performance optimization depends on the analysis of the modal power distribution and the modal relative-phase distribution.
The power distribution directly affects the beam quality of fiber \cite{1605337}, constrains the power-transmission efficiency of the waveguide \cite{Ohkubo_Kubo_Idei_Sato_Shimozuma_Takita_1997}, and serves as a key metric for evaluating mode-coupling efficiency \cite{Oda01042012} and validating mode-conversion performance \cite{8642538}.
The modal relative-phase distributions among different modes can induce interference effects, resulting in power loss \cite{5597962}, mode mixing \cite{Ramachandran2008}, and coupling loss \cite{10221692}.
Parasitic modes from assembly errors and device coupling reduce mode conversion efficiency in microwave systems\cite{10254364}. Amplitude and phase mismatch across channels causes instability in power combination systems by exciting low-order modes in oversized waveguide\cite{10439262}.

Existing modal analysis techniques, however, face a trade-off between accuracy, complexity, and measurement intrusiveness.
The methods fall into three principal categories: invasive waveguide measurement methods, post–mode-separation analysis methods, and radiation-field inversion methods.

Invasive waveguide measurement methods place sensors inside the waveguide to determine the modal power distribution, primarily via electric and magnetic probes \cite{1126264} or dual-antenna frequency response \cite{1256771} to measure. However, these methods are constrained by mode-amplitude stability and antenna spacing.
Post–mode‑separation analysis methods extract modal information via mode‑separation techniques and then analyze the power and phase of each mode. In oversized waveguides, modes are separated by openings in the waveguide wall for subsequent analysis \cite{1125448}. 
In multimode fibers and photonic integrated circuits, optical components such as directional couplers \cite{Dai:13} are used, often combined with selective phase matching \cite{Uden2014U} and mode-spectrum mapping \cite{Zhou:18} to extract modal information.
Both approaches modify the physical waveguide structure, inducing mode conversion, coupling effects, and even excitation of higher-order modes, which in turn compromise the measurement accuracy.

Radiation-field inversion methods infer the internal modal content by measuring the external radiation field of the waveguide.
This category primarily encompasses thermographic imaging \cite{Wang20038mmTM} and infrared-camera-based intensity-pattern retrieval \cite{OdaKTs2010}, but both yield relatively low accuracy.
Although the study by  \cite{Manuylovich_Dvoyrin_Turitsyn_202010.5555/3495724.3495945} proposed a linear method based on matrix operations for optical fiber mode decomposition, this approach cannot be directly transferred to waveguide analysis. The fundamental reason lies in the fact that the AFs and FFs of waveguide modes contain zero components, which can render the inversion matrix ill-conditioned, thereby preventing the acquisition of a stable numerical solution.
\cite{Liu:18,9906576} employ deep neural networks to predict mode power distribution and mode coefficients from light intensity or electric field inputs. However, these approaches suffer from limited accuracy and require extensive datasets for generalization.
\cite{Lu:13} utilizes a stochastic parallel gradient descent (SPGD) method using FF data for non-convex optimization, which achieves high computational efficiency but suffers from twin-image ambiguity, making it difficult to obtain a unique solution. Its improved hybrid optimization framework \cite{Li:17} first employs the Genetic Algorithm (GA) for the non-convex optimization of both near field(NF) and FF data. It then focuses on the NF measurement using the SPGD for local search. This approach achieves high accuracy, but the global optimization process incurs high computational costs. 
This strategy is effective in fiber systems because the NF is obtained via a 4-f system to indirectly characterize the AF, avoiding direct interference. However, in open microwave waveguides, AF measurement directly alters the electromagnetic geometry of the waveguides, resulting in unintended field distributions.
Vector network analyzer is used to measure the attenuation of the parasitic mode \cite{10254364} and retrieve the modal power distribution with high accuracy \cite{6186834}, but the complex equipment limits its application.
Especially for high-power waveguides, equivalent tests are typically conducted on low-power equipment to prevent instrument damage \cite{amech2005}, which further degrades accuracy.
Radiation-field inversion methods struggle to balance accuracy, equipment complexity, and computational cost.
Essentially, a robust, accurate and non-invasive method for retrieving complete modal information (power and phase) remains a major challenge.

Viewing this challenge through the formal lens of a phase retrieval problem provides deeper insights.  
Recovering the complete modal information, particularly the MRPD, from FF amplitudes is the central objective of this task, precisely because the process of obtaining the MPD is linear.
While classic algorithms like Gerchberg-Saxton (GS) exist \cite{1971Phase, Gerchberg_1972}, they are notoriously sensitive to initialization and prone to stagnating in local minima \cite{Wackerman:91,Lu:93,Takajo:97}.
More recent, powerful frameworks such as PhaseLift \cite{Candès21432}, PhaseCut \cite{10.1007/s10101222}, and Wirtinger Flow (WF) \cite{Candès_Li_Soltanolkotabi_2015} have demonstrated remarkable success.
However, their theoretical guarantees of convergence all hinge on a critical, shared assumption: the sampling vectors are drawn from a random distribution (e.g., Gaussian) \cite{Candès_Li_Soltanolkotabi_2015,GAO202195,CANDES2015277}.
This randomness endows the loss landscape with favorable geometric properties, enabling reliable convergence.

 In waveguide modal analysis, however, this fundamental assumption is violated.  The sampling vectors are not random;  they are deterministically defined by the physical eigenmodes of the waveguide structure.  This deterministic nature invalidates the theoretical underpinnings and initialization strategies of conventional phase retrieval methods, exposing a critical research gap and explaining why a direct application of these tools often fails in this context. 

To solve this deterministic, nonconvex optimization problem, we turn to a different paradigm, inspired by the training of large-scale neural networks.
Recent studies have revealed that employing a large-step-size optimization strategy can induce an ``implicit regularization'' effect, guiding the optimizer to escape sharp, poor local minima and converge to flatter, more robust solutions \cite{10.5555/3495724.3495945, pmlr-v202-mohtashami23a, wang2023good}.
This approach offers a powerful mechanism to overcome the stagnation issues that plague traditional gradient-based methods, especially when a good initialization is not available, without relying on the statistical properties of random sampling.
Inspired by this, we introduce a novel framework that, for the first time, applies this large-step-size principle to the deterministic phase retrieval problem in waveguide analysis.  By formulating the task as a nonconvex optimization problem under a power-normalization constraint and utilizing the AdaMax optimizer with a large-step strategy, we can robustly recover both modal power and phase coefficients. This work contributes as follows:
\begin{enumerate}
\IEEEpubidadjcol
\item \textbf{A New Optimization Paradigm for Deterministic Inverse Problems:} We are the first to apply the large-step optimization paradigm, inspired by deep learning theory, to the deterministic phase retrieval problem in waveguide modal analysis. 
\item \textbf{Systematic Analysis of Twin-Image Ambiguity:} The aperture radiation model, as an effective Fourier transform, introduces distinct twin solutions. The joint optimization using sampling information generated from the true solution makes the unique determination of the candidate solutions possible.
\item \textbf{Theoretical Foundation and Objective Function Design:} Our objective function incorporates not only the primary optimization goal but also a power normalization constraint. To accommodate large-step optimization, a high-order objective formulation is adopted, and the theoretical equivalence between parameter gradient descent and Wirtinger gradient descent is rigorously established, thereby validating the rationality of our objective design and the effectiveness of its task representation.
\item \textbf{Comprehensive Evaluation and Comparative Analysis:} We first visualized the underlying principles of the proposed method through its loss landscape, providing intuitive insight into the optimization dynamics. Then, under specified experimental conditions, we conducted systematic evaluations to verify its scalability in handling a large number of modes, its robustness under various SNR levels, and the effectiveness of sample density in suppressing noise. Finally, a detailed comparison with other state-of-the-art approaches demonstrated substantial improvements in both accuracy and computational efficiency.

\end{enumerate}

\section{Theoretical Modeling}\label{sec:theo_model}
\subsection{Modal Analysis Framework}\label{subsec:modal_analysis_framework}
This study involves sampling the amplitude of the AF and FF. On the waveguide AF (in cylindrical coordinates with the aperture center as the origin and the $z$-axis aligned with the waveguide axis)
, we sample the electric field amplitudes in the $\rho$ (radial)
 and $\varphi$ (azimuthal) directions, obtaining $M_\text{AF}$ samples in total. Similarly, on the FF hemispherical surface (in spherical coordinates sharing the same origin and $z$-axis as the AF system)
, we sample the electric field amplitudes in the $\theta$ (polar)
and $\varphi$ (azimuthal) directions, acquiring $M_\text{FF}$ samples. The total sample size is $M = M_\text{AF} + M_\text{FF}$. Assuming there are N propagating modes in the waveguide, we compute the power-normalized field strength amplitudes corresponding to different modes at each sampling point based on the theoretical formulas or simulation results provided in \cref{subsec: wg_formula}. The function values of all modes at the $i$-th sampling point form a sampling vector $\boldsymbol{u}_i = \bigl[u_1, u_2, \dots, u_N \bigr]^\top \in \mathbb{C}^{N}$. Accordingly, the modal analysis problem can be formulated as solving
\begin{equation}\label{eq:problem}
  y_i=\langle \boldsymbol{u}_i,\boldsymbol{z}\rangle
^2, i=1,2,\dots M .
\end{equation}
Here, $\boldsymbol{z} = [z_1,z_2,\cdots,z_N]^\top\in \mathbb{C}^{N}$ is the complex modal-coefficient vector: the moduli $|z_i|$ of its components equal the square roots of the modal power distribution, and the arguments $\arg (z_i)$ reflect the modes' relative-phase distribution.

\subsection{Waveguide Formulation}\label{subsec: wg_formula}

This study selects a circular waveguide as the test case.
The higher-order modes in the AF are excited by the altered electromagnetic geometry during measurement and are simulated using a noise model. The reflection and diffraction from the waveguide aperture are neglected, and the FF is computed via the aperture radiation model.
Based on \cite{5265485,doi:10.1049/PBEW019E_ch7,doi:10.1049/PBEW019E_ch10,pozar2011ch3}, we derive analytical expressions for the modal propagation power and FF intensity of each mode, and present them in \cref{table:circular_far_power}.
In the table, $(r,\theta,\varphi)$ denotes the coordinates with respect to the aperture centre; the permittivity $\varepsilon$ and the permeability $\mu$ are taken as the vacuum values $\varepsilon_0$ and $\mu_0$ (in both the waveguide and the radiation region). $\lambda$ is the wavelength of the mode; $k$ is its wave number; $k_{c,nm}$ is the cutoff wave number; and $\epsilon_{0n}$ is the Neumann factor \cite{5265485}:
\begin{equation}
  \epsilon_{0n} = \begin{cases} 1, & n=0 \\2, & n>0 \end{cases}.
\end{equation}
To ensure consistency between the magnitudes of the electric and magnetic fields, it is recommended that the axial field amplitudes of the TE and TM modes satisfy $A_{nm} = B_{nm} / \eta$, where $\eta = \sqrt{\mu/\varepsilon}$ denotes the wave impedance inside the waveguide.

\begin{table*}[!tbp]
    \centering
    \caption{Analytical Expressions for Far-Field Polarized Intensity Components and Modal Propagation Power of Circular Waveguide TE/TM Modes}
    \label{table:circular_far_power}
    \begin{tabular}{ c c c }
    \toprule 
    \textbf{Field \& Power}& \textbf{TE Mode}  & \textbf{TM Mode}
    \\ \midrule 
    $H_{\text{AF},nm,z}^{\text{TE}} $, $E_{\text{AF},nm,z}^{\text{TM}}$ &
    $A_{nm}e^{-j \phi_{nm}^\text{TE}} J_n{\left (k_{c,nm}'\rho \right )} \begin{cases}\cos n\varphi\\\sin n\varphi\end{cases}$ 
    & $B_{nm}e^{-j \phi_{nm}^\text{TM}} J_n{\left (k_{c,nm}\rho\right )} \begin{cases}\cos n\varphi\\\sin n\varphi\end{cases}$ 
    \\
    $E_{\text{AF},nm,\rho}^{\text{TE}}$, $E_{\text{AF},nm,\rho}^{\text{TM}}$  &
    $-j\dfrac{n \omega \mu}{\rho k_{c,nm}'^2}A_{nm} J_n \left(k_{c,nm}'\rho\right)e^{-j\beta_{nm}z}\begin{cases} -\sin  n\varphi\\\cos n\varphi\end{cases}$ &
    $-j\dfrac{\beta_{nm}}{k_{c,nm}}B_{nm} J_n' \left(k_{c,nm}\rho\right)e^{-j\beta_{nm}z}\begin{cases}\cos n\varphi\\\sin n\varphi\end{cases}$
    \\
    $E_{\text{AF},nm,\varphi}^{\text{TE}}$, $E_{\text{AF},nm,\varphi}^{\text{TM}}$  &
    $j\dfrac{\omega \mu }{k_{c,nm}'}A_{nm} J'_n \left(k_{c,nm}'\rho\right)e^{-j\beta_{nm}z}\begin{cases}\cos n\varphi\\\sin n\varphi\end{cases}$ &
    $-j\dfrac{n\beta_{nm} }{\rho k_{c,nm}^2}B_{nm} J_n \left(k_{c,nm}\rho\right)e^{-j\beta_{nm}z}\begin{cases} -\sin  n\varphi\\\cos n\varphi\end{cases}$
    \\ 
    $E_{\text{FF},nm,\theta}^{\text{TE}}$, $E_{\text{FF},nm,\theta}^{\text{TM}}$  &
    $\begin{aligned}[t] 
      &- \frac{A_{nm}e^{-j \phi_{nm}^\text{TE}}\omega \mu}{k_{c,nm}'^2} \frac{n }{2 r} \left (1 + \frac{\beta_{nm}}{k} \cos \theta \right )  \\
      &\times \frac{J_n(k_{c,nm}' a)J_n(k a \sin \theta)}{\sin \theta} e^{-j [kr-(n+1)\frac{\pi}{2}]} \begin{cases} -\sin  n\varphi\\\cos n\varphi\end{cases} 
    \end{aligned}$
    & $\begin{aligned}[t] 
      &- \frac{B_{nm}e^{-j \phi_{nm}^\text{TM}} }{k_{c,nm}\sin \theta }
      \frac{ka}{2r}
      \left (\frac{\beta_{nm}}{k} + \cos \theta \right ) \\
      &\times\frac{J_n(k a \sin \theta) J'_n(k_{c,nm} a)}{1 - \left (\frac{k_{c,nm}}{k \sin \theta}\right )^2} e^{-j [kr-(n+1)\frac{\pi}{2}]}\begin{cases}\cos n\varphi\\\sin n\varphi\end{cases} 
      \end{aligned}$
    \\ 
    $E_{\text{FF},nm,\varphi}^{\text{TE}}$, $E_{\text{FF},nm,\varphi}^{\text{TM}}$ &
    $\begin{aligned}[t] 
    &  \frac{A_{nm}e^{-j \phi_{nm}^\text{TE}}\omega \mu}{k_{c,nm}^2} \frac{k a }{2 r} \left (\frac{\beta_{nm}}{k} + \cos \theta \right ) \\
    &\times\frac{J_n(k_{c,nm}' a) J'_n(k a \sin \theta)}{1-\left (\frac{k \sin \theta}{k_{c,nm}}\right )^2}  e^{-j [kr- (n+1)\frac{\pi}{2}]} \begin{cases}\cos n\varphi\\\sin n\varphi\end{cases} 
    \end{aligned}$
    & $0$
    \\  
    $P_{nm}^\text{TE}$, $P_{nm}^\text{TM}$ &
    $\dfrac{ \pi \omega \mu \beta_{n,m} |A_{nm}|^2 }{2k_{c,nm}'^4 \epsilon_{0 n}}(\chi_{nm}'^2-n^2)J_n ^2(\chi_{nm}')$
    &
    $\dfrac{\pi \omega \varepsilon\beta_{nm} |B_{nm}|^2}{2k_{c,nm}^4 \epsilon_{0n}}\chi_{nm}^2 \left[J_n' (\chi_{nm})\right]^2$
    
    \\
    \bottomrule 
    \end{tabular}
  \end{table*}

\subsection{Power-Normalization Constraint}\label{subsec:normalized}
For the $N$ propagation modes in the waveguide (including $\mathrm{TE}$ and $\mathrm{TM}$ modes), $A_{nm}e^{-j \phi_{nm}^\mathrm{TE}}$ and $B_{nm}e^{-j \phi_{nm}^\mathrm{TM}}$ are separated from the AF and FF intensity functions of the modes, yielding
\begin{equation}
  E_{s,i,\alpha}=A_ie^{-j\phi_i}v_{s,i,\alpha},~\alpha\in\begin{cases}\{\rho,\varphi\},~s=\mathrm{AF}\\\{\theta,\varphi\},~s=\mathrm{FF}\end{cases}.
\end{equation}
Here, $v_{\text{AF},i,\rho},~v_{\text{AF},i,\varphi},~v_{\text{FF},i,\theta}$ and $v_{\text{FF},i,\varphi}$ are the AF and FF radiation-field-strength functions excluding the axial amplitudes $A_{nm}$ and $B_{nm}$, as listed in \cref{table:circular_far_power}.
The radiation power of a single mode is defined as
\begin{equation}
\begin{aligned}
    P_{\mathrm{\mathrm{rad}},i} &=A_i^2
    \iint_\Sigma \frac{|v_{\text{AF},i,\rho}|^2 +|v_{\text{AF},i,\varphi}|^2 }{2\eta} \mathrm{d}S \\
    &=A_i^2\iint_\Omega \frac{|v_{\text{FF},i,\theta}|^2 +|v_{\text{FF},i,\varphi}|^2 }{2\eta} \mathrm{d}S 
    \equiv A_i^2 p_{\mathrm{\mathrm{rad}},i}.
\end{aligned}
\end{equation}
Here, $\Sigma$ denotes the aperture surface in the AF, 
and $\Omega$ represents the hemispherical surface in the FF.  $\eta$ is the wave impedance of the radiation region (for which the vacuum impedance $\eta_0$ is used in this paper). The modal radiation power $p_{\mathrm{rad},i}$ is evaluated under the normalization $A_{nm}=B_{nm}/\eta=1$.
The total FF radiation power $P_{\mathrm{rad}}$ satisfies energy conservation,
\begin{equation}
  P_{\mathrm{rad}} = \sum_i^N P_{\mathrm{rad},i} = \sum_i^N A_i^2p_{\mathrm{rad},i}.
\end{equation}
The power ratio of each mode is defined as
\begin{equation}
  d_i =  \frac{A_i^2p_{\mathrm{rad},i}}{P_{\mathrm{\mathrm{rad}}}}.
\end{equation}
In waveguide structures, energy loss is small, and the power at the waveguide aperture is predominantly transmitted through FF radiation. Therefore, it can be approximated that
\begin{equation}
  \frac{p_i}{P}\approx \frac{p_{\mathrm{rad},i}}{P_{\mathrm{rad}}},
\end{equation}
where $P$ denotes the power at the waveguide aperture, which needs to be measured in practice, and $p_i$ denotes the modal propagation power when $A_{nm} = B_{nm} / \eta = 1$.
The normalized FF intensity functions for each mode are introduced as
\begin{equation}
u_{s,i,\alpha}=v_{s,i,\alpha}\sqrt{P/p_i},~\alpha\in\begin{cases}\{\rho,\varphi\},~s=\mathrm{AF}\\\{\theta,\varphi\},~s=\mathrm{FF}\end{cases}.
\end{equation}
The total field can be expressed as
\begin{equation}\label{eq:sum}
   E_{s,\alpha}=\sum_i z_i u_{s,i,\alpha},~\alpha\in\begin{cases}\{\rho,\varphi\},~s=\mathrm{AF}\\\{\theta,\varphi\},~s=\mathrm{FF}\end{cases}.
\end{equation}
The complex coefficients are given by
\begin{equation}\label{eq:coeffs}
  z_i = |z_i| e^{-j \phi_i } = \sqrt{d_i} e^{-j \phi_i},
\end{equation}
and satisfy the power-normalization constraint
\begin{equation}\label{eq:reg}
  \sum_i d_i = \sum_i |z_i| ^2 = \sum_iz_i^*z_i =  1  .
\end{equation}

\subsection{Twin-Image Ambiguity}\label{subsec:twin-image-ambiguity}

\begin{figure}[!tbp]
    \centering
    \includegraphics[width=0.8\linewidth]{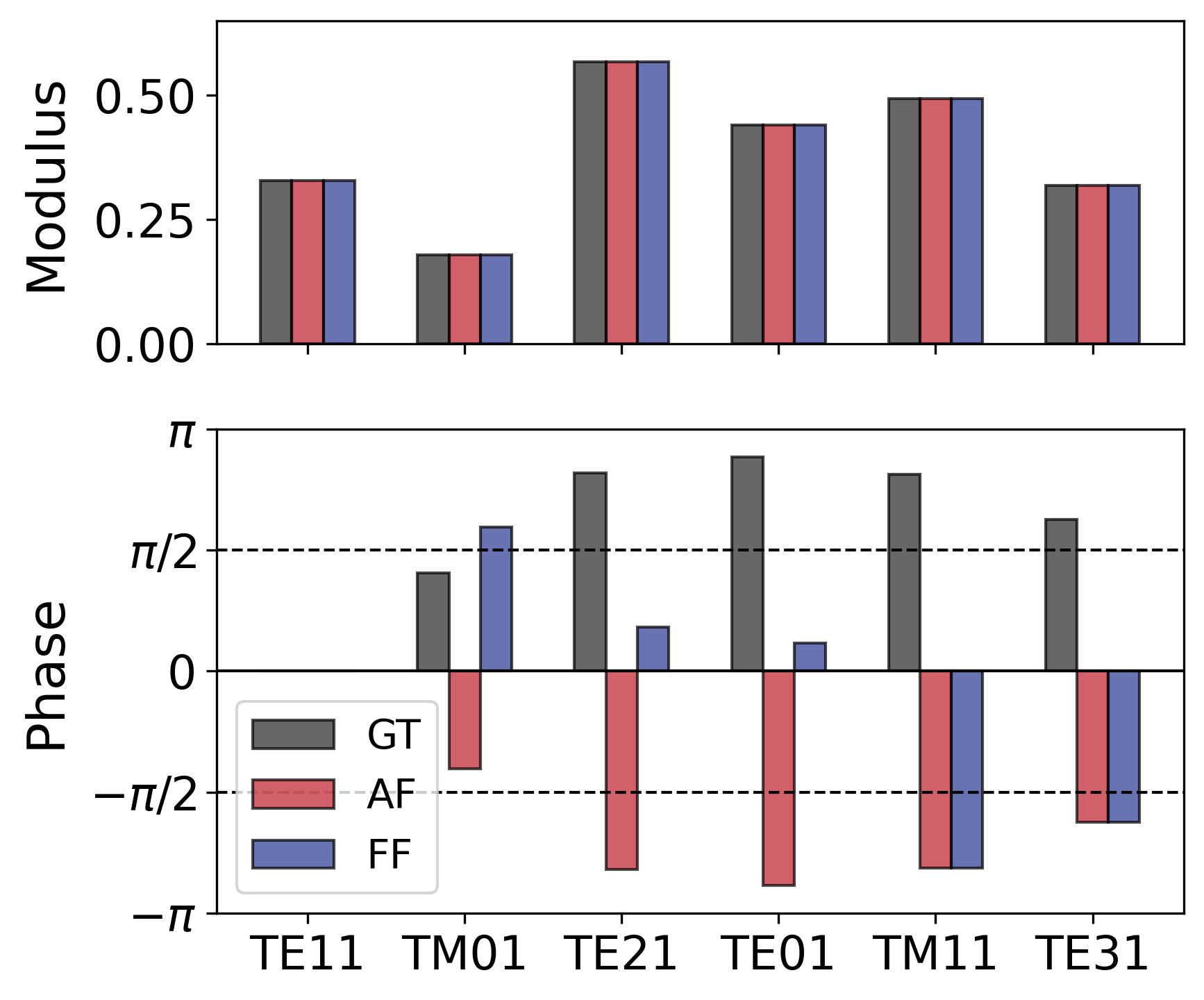}
    \caption{Schematic diagram of phase ambiguity induced by twin-image ambiguity. The legend indicates the true complex coefficients (GT, Ground Truth) and their twin counterparts generated in the AF and the FF. The upper plot demonstrates that the modulus remains unaffected by the twin images; the lower plot, using the TE11 mode as the phase reference, reveals that the some twin phases in AF and FF are symmetric about the $\pm \pi / 2$ axis relative to the GT phase.}
    \label{fig:ambiguity}
\end{figure}

In fiber modal analysis, a single set of amplitude data from the AF and the FF may correspond to two distinct relative phase distributions. This phenomenon is referred to as the ``twin image'' problem in phase retrieval. This study defines the phase distribution that yields the same amplitude image as the true phase as the twin phase. It is noted that the AF and the FF each correspond to a different twin phase.
In the phase retrieval of Fourier sampling vectors, the phase distribution corresponding to this phenomenon can be obtained by conjugate inversion, which constitutes a form of trivial ambiguity \cite{Fannjiang_2012}.
Fiber mode analysis based on NF and FF measurements encounters this class of problems \cite{Manuylovich_Dvoyrin_Turitsyn_202010.5555/3495724.3495945}. Similarly, phase retrieval for the waveguide AF faces an identical issue.

As noted in \cite{Manuylovich_Donodin_Turitsyn_2021}, the transformation from the near field to the FF in optical fibers involves a Fourier transform. This results in distinct twin phases between the NF and FF. The aperture radiation model for waveguides essentially corresponds to an equivalent Fourier transform, thereby enabling the recovery of the true relative phase distribution of fiber modes.  However, in waveguide mode analysis, the presence of null field components for certain modes in specific directions prevents the calculation of the correct $\boldsymbol{z}$ vector using the least-squares method (or pseudo-inverse solution) proposed in \cite{Manuylovich_Donodin_Turitsyn_2021}. Consequently, the subsequent nonlinear phase screening process becomes infeasible. This renders the aforementioned method impractical for waveguide mode analysis. We attribute this fundamental difference to the well-defined linear polarization characteristics inherent to fiber Linearly Polarized (LP) modes, which are generally not possessed by waveguide modes.

The AF and FF of the waveguide can be decomposed into amplitude and phase components, with the amplitude described by real-valued function and the phase represented by a complex exponential term.
For $N$ modes (including TE and TM modes), the power normalized AF and FF function, referring to \cref{table:circular_far_power}, can be written as
\begin{equation}
    \begin{cases}
        u_{\text{AF},i}(\rho,\varphi) = |u_{\text{AF},i}|e^{-j\psi_{\text{AF},i}}\\
        u_{\text{FF},i}(r,\theta, \varphi) = |u_{\text{FF},i}|e^{-j(kr+\psi_{\text{FF},i})}
    \end{cases},
\end{equation}
where $|u_{i}|$ represents the amplitude component and $\psi_{i}$ depends on the phase information of each mode. 
The total field can be expressed as
\begin{equation}\label{eq:AF_FF_sum_ambiguity}
        \begin{cases}
            E_\text{AF}
        = \sum_i^N |z_i| |u_{\text{AF},i}|e^{-j(\psi_{\text{AF},i}+\phi_{i})}
            \\
            E_\text{FF} 
            = \sum_i^N |z_i| |u_{\text{AF},i}|e^{-j(kr+\psi_{\text{FF},i}+\phi_{i})}
        \end{cases},
\end{equation}
In practical experiments, only the amplitudes of AF and FF fields are available, and the modulus operation exhibits conjugate symmetry, i.e., $|E| = |\overline{E}|$.
Conjugate inversion leads to the so-called ``twin imag'' phenomenon, resulting in the ambiguity of the solution.
According to \cref{eq:AF_FF_sum_ambiguity}, the twin complex coefficients for the AF and the FF can be readily obtained.
\begin{equation}
    \begin{cases}
        z_{\text{AF},i}' = |z_i|e^{j(\psi_{\text{AF},i}+\phi_{i})}\\
        z_{\text{FF},i}' = |z_i|e^{-j[-2kr-(2\psi_{\text{FF},i}+\phi_{i})]}
    \end{cases},
\end{equation}
According to \cref{table:circular_far_power}, in the circular waveguide, $\psi_{\text{AF},i} = 0,~ \psi_{\text{FF},i} = (n+1)\frac{\pi}{2}$. Taking a circular waveguide supporting the six modes TE11, TM01, TE21, TE01, TM11, and TE31 as an example, \cref{fig:ambiguity} illustrates the corresponding modal power distribution and relative modal phase distribution in both the AF and the FF.

\subsection{Noise Modeling }\label{subsec: noise}

The noise field is non-stationary and can, over an infinitely long time, be decomposed into a superposition of orthogonal harmonic components.
Assuming the medium is frequency-nondispersive, the sum of the time-averaged energy flux densities in all directions is
\begin{equation}
  \overline{S}_\text{noise} = \frac{\mathrm{Var}(E_{\text{noise}})}{2\eta} = \frac{\sigma^2}{2\eta}.
\end{equation}
On the sampling hemisphere, if the noise power is uniformly distributed, then
\begin{equation}
  P_\mathrm{noise,AF} = \pi a^2 \overline{S}_\mathrm{noise,AF},~P_\mathrm{noise,FF} = 4\pi r^2 \overline{S}_\mathrm{noise,FF}.
\end{equation}
Here, $a$ denotes the radius of the waveguide, and $r$ represents the distance from the aperture center (the origin) to the FF sample point.
Let the signal power be the radiated power $P_\mathrm{rad}$, and define the SNR as
\begin{equation}\label{eq:SNR}
  \mathrm{SNR} = 10\log_{10} \left ( \frac{P_\mathrm{rad}}{P_{\text{noise}}} \right ).
\end{equation}
The electric field noise is typically modeled as additive zero-mean complex circularly symmetric Gaussian noise, whose real and imaginary parts are independent and identically distributed Gaussian random variables. Consequently, its amplitude follows a Rayleigh distribution. Since the polarization components are orthogonal to each other, the noise in different polarization components can be simplified as independent Gaussian random variables. 
Thus, in both AF and FF measurements, the polarization components of the electric field are corrupted by additive Gaussian noise. Both measurements include background noise $n_{\mathrm{FF}}$, while the AF measurement additionally contains a noise term $n_{\mathrm{AF}}$ approximating the unintended fields induced by the probe, i.e., perturbations of the waveguide’s electromagnetic geometry (with the aperture as reference, hence denoted by the AF subscript). This noise model can be expressed as:

\begin{equation}
    \begin{cases}
        |E_{\text{AF},\rho,i}| \gets |E_{\text{AF},\rho,i}| + n_{\text{AF},\rho,i} + n_{\text{FF},\rho,i}\\
        |E_{\text{AF},\varphi,i}| \gets |E_{\text{AF},\varphi,i}| + n_{\text{AF},\varphi,i} + n_{\text{FF},\varphi,i}\\
        |E_{\text{FF},\theta,i}| \gets |E_{\text{FF},\theta,i}| + n_{\text{FF},\theta,i}\\
        |E_{\text{FF},\varphi,i}| \gets |E_{\text{FF},\varphi,i}| + n_{\text{FF},\varphi,i}  
    \end{cases},
\end{equation}
where $n_{\text{AF},\rho,i},\, n_{\text{AF},\varphi,i} \sim \mathcal{N}(0, \sigma_{\text{AF}}^2/2)$ and $n_{\text{FF},\theta,i},\, n_{\text{FF},\varphi,i} \sim \mathcal{N}(0, \sigma_{\text{FF}}^2/2)$ represent the noise components in the respective field components. The standard deviations $\sigma_{\text{AF}}$ and $\sigma_{\text{FF}}$ are given by the following equations:
\begin{equation}
    \sigma_{\text{AF}} = \sqrt{2\eta\overline{S}_\text{noise,AF}}, ~ \sigma_{\text{FF}} = \sqrt{2\eta\overline{S}_\text{noise,FF}}.
\end{equation}

\begin{figure*}[!b]
    \centering
    \includegraphics[width=0.8\linewidth]{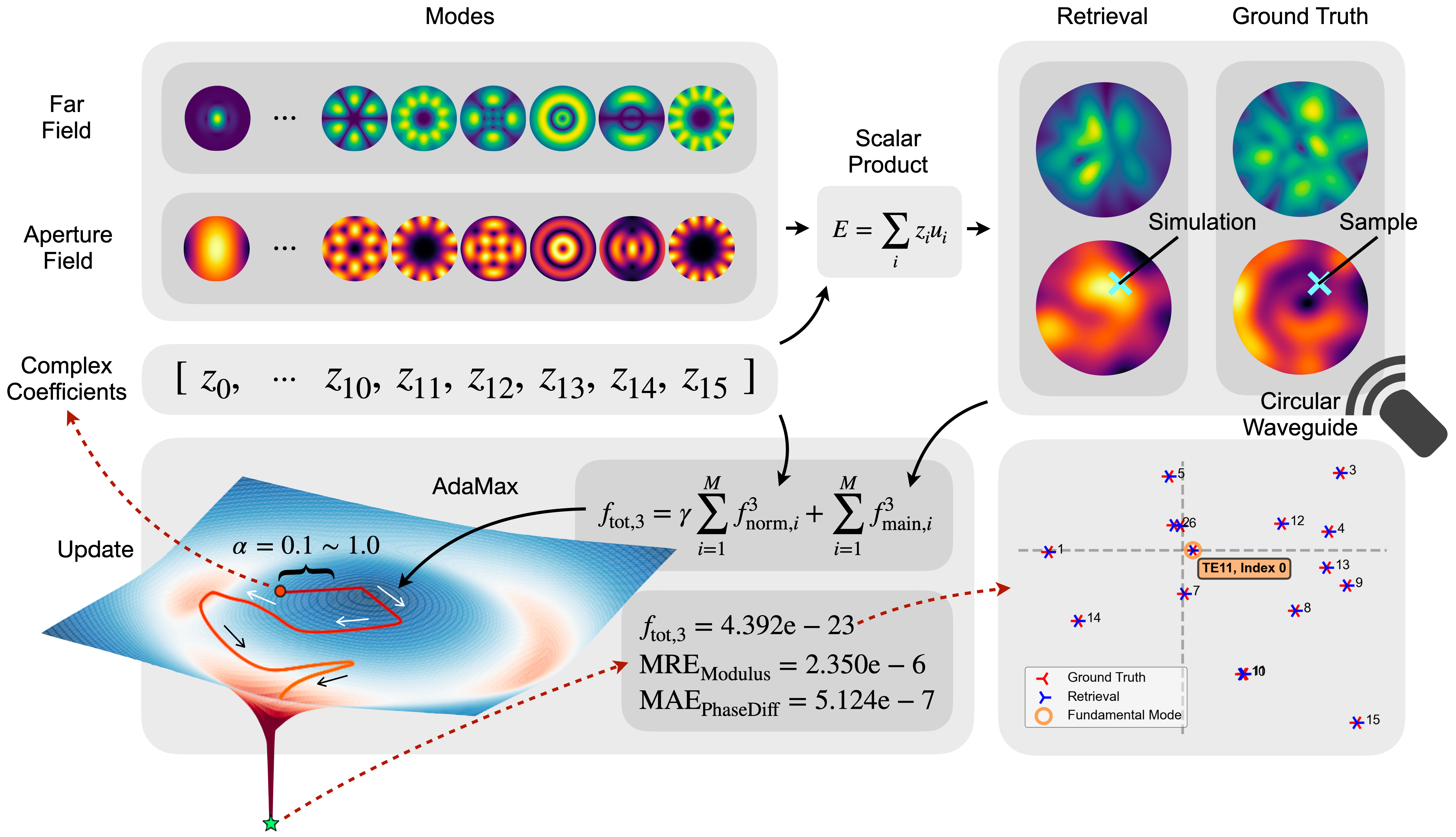}
    \caption{Schematic diagram of the modal analysis method based on large-step nonconvex optimization. The workflow begins with acquiring amplitude-only measurements from the AF and FF, which serve as labels, and obtaining the field distributions for all potential modes through simulation. The unknown complex coefficients are randomly initialized and then iteratively refined in a nonconvex optimization loop. In each iteration, the current coefficients are used to calculate two loss components: a main loss ($f_{\mathrm{main}}^3$) based on the difference between the measured and simulated amplitudes, and a power normalization loss ($f_{\mathrm{norm}}^3$). An advanced optimizer with a large-step-size strategy minimizes the total loss ($f_{\mathrm{tot},3}$) until the coefficients converge to their true values. Black arrows indicate the workflow process, while red arrows represent explanatory relationships or updates between modules.}
    \label{fig:overview}
\end{figure*}

\section{Proposed Method}\label{sec:method}
\subsection{Method Overview}\label{subsec:overview}

The core challenge of mode analysis is to accurately determine the complex coefficients (both amplitude and phase) of all propagating modes from limited measurement data. 
As described in \cref{subsec:twin-image-ambiguity}, recovering phase from amplitudes alone involves inherent ambiguity. The AF and FF each correspond to a pair of twin phases (see \cref{fig:ambiguity}), with the true phase lying at the intersection of the two pairs. As noted in \cref{section:introduction}, AF measurement alters the electromagnetic geometry of the waveguides, introducing unintended field distributions, while relying solely on FF amplitudes cannot resolve the twin-image ambiguity. Therefore, we use AF amplitudes as auxiliary information. Unlike \cite{Li:17}, which employs NF amplitudes for non-convex optimization in the second stage, we adopt a more concise joint optimization strategy.
We propose a robust method based on large-step-size nonconvex optimization, which reconstructs these coefficients using only amplitude measurements from the AF and FF. The overall workflow of our proposed method is illustrated in \cref{fig:overview}. The process begins with two sets of inputs: a series of amplitude measurements acquired at various sampling points in the AF and FF, which serve as the ground truth labels, and a basis set of all possible modal field distributions, obtained a priori via numerical simulation or analytical calculation. The unknown complex coefficients are first randomly initialized and normalized. Within the optimization loop, these coefficients are used to compute a total loss function, $f_{\mathrm{tot},3}$, which consists of two key components. The main loss, $f_{\mathrm{main}}^3$, quantifies the discrepancy between the measured amplitude labels and the simulated amplitudes, which are synthesized by a linear combination of the modal basis and the current coefficient estimates. The second component, a power normalization loss $f_{\mathrm{norm}}^3$, acts as a physical constraint. Finally, an advanced optimizer is employed with a large-step-size strategy to efficiently navigate the nonconvex loss landscape. The optimizer iteratively updates the complex coefficients to minimize the total loss until a predefined convergence criterion is met.

In following subsections, we elaborate on the theoretical foundation and mathematical formulation of the proposed method.
\subsection{Wirtinger Gradient Descent Framework}\label{subsec:Wirtinger}

In the complex space $\mathbb{C}^N \simeq \mathbb{R}^{2N}$, any complex vector can be written as 
\begin{equation}
  \boldsymbol{z}=\boldsymbol{x}+i\boldsymbol{y}\quad  (\boldsymbol{x},\,\boldsymbol{y}\in\mathbb{R}^N).
\end{equation}
The Wirtinger derivatives are defined as
\begin{equation}
  \frac{\partial}{\partial\boldsymbol{z}}=\frac{1}{2}\left(\frac{\partial}{\partial\boldsymbol{x}}-j\frac{\partial}{\partial\boldsymbol{y}}\right),\quad \frac{\partial}{\partial\overline{\boldsymbol{z}}}=\frac{1}{2}\left(\frac{\partial}{\partial\boldsymbol{x}}+j\frac{\partial}{\partial\boldsymbol{y}}\right),
\end{equation}
where $\boldsymbol{z}$ and its conjugate $\overline{\boldsymbol{z}}$ are treated as independent variables, as their derivatives satisfy mutual independence:
\begin{equation}
  \frac{\partial\overline{\boldsymbol{z}}}{\partial\boldsymbol{z}}= \boldsymbol{0},\quad \frac{\partial\boldsymbol{z}}{\partial\overline{\boldsymbol{z}}}=\boldsymbol{0}.
\end{equation}
Therefore, we can construct a conjugate coordinate system as
\begin{equation}
  \begin{bmatrix} \boldsymbol{z} \\ \overline{\boldsymbol{z}} \end{bmatrix}\in \mathbb{C}^N \times \mathbb{C}^N.
\end{equation}
In this coordinate system, for a small perturbation \(\Delta\mathbf{z}\in\mathbb{C}^N\), the Taylor expansion of \(f(\mathbf{z}+\Delta\mathbf{z})\) about \(\mathbf{z}\) is given by:
\begin{equation}
  \begin{aligned}
  f(\boldsymbol{z}+\Delta\boldsymbol{z})&= f(\boldsymbol{z})+\left (\nabla_{c}f(\boldsymbol{z})\right )^{*}\begin{bmatrix}\Delta\boldsymbol{z}\\\Delta\overline{\boldsymbol{z}}\end{bmatrix}\\
  &+\frac{1}{2}{\begin{bmatrix}\Delta\boldsymbol{z}\\\Delta\overline{\boldsymbol{z}}\end{bmatrix}}^{*}\nabla_c^{2}f(\boldsymbol{z}){\begin{bmatrix}\Delta\boldsymbol{z}\\\Delta\overline{\boldsymbol{z}}\end{bmatrix}} + o(|\Delta \boldsymbol{z}|^2)
  \end{aligned},
\end{equation}
where $\nabla_c f(\boldsymbol{z})$ and $\nabla_c^2f(\boldsymbol{z})$ are the Wirtinger-based gradient and Hessian in the conjugate coordinate system.
The conjugate‑coordinate update rule reads
\begin{equation}
\begin{bmatrix}\boldsymbol{z}_{t+1}\\\bar{\boldsymbol{z}}_{t+1}\end{bmatrix}=\begin{bmatrix}\boldsymbol{z}_t\\\bar{\boldsymbol{z}}_t\end{bmatrix}-2\alpha\begin{bmatrix}\left(\partial f/\partial\boldsymbol{z}\right)^*|_{\boldsymbol{z}=\boldsymbol{z}_t}\\\left(\partial f/\partial\bar{\boldsymbol{z}}\right)^*|_{\boldsymbol{z}=\boldsymbol{z}_t}\end{bmatrix}.
\end{equation}
For $f \in C^1(\mathbb{C}^N, \mathbb{R})$, there is conjugate symmetry
\begin{equation}
  \overline{\frac{\partial f}{\partial\boldsymbol{z}}}=\frac{\partial f}{\partial\bar{\boldsymbol{z}}},
\end{equation}
so we can derive the update rule for $\boldsymbol{z}$:
\begin{equation}
  \begin{aligned}
    \left(\partial f/\partial\boldsymbol{z}\right)^*&= \frac{1}{2}\left ({\partial f}/{\partial\boldsymbol{x}}^\top + j{\partial f}/{\partial\boldsymbol{y}}^\top\right ) \\
    &= \frac{1}{2}\left (\nabla_{\boldsymbol{x}} f + j\nabla_{\boldsymbol{y}} f\right ).
  \end{aligned}
\end{equation}
If the real part $\boldsymbol{x}$ and the imaginary part $\boldsymbol{y}$ are treated as parameters $\theta$, then the above equation is equivalent to
\begin{equation}
  \theta_{t+1} = \theta_t-\alpha\nabla_{\theta}f|_{\theta=\theta_t}.
\end{equation}
Here, $\alpha$ is the update step size (i.e., the learning rate in deep learning).
This demonstrates that Wirtinger gradient descent is mathematically equivalent to standard gradient descent on the real parameter vector, and hence can be implemented directly via automatic differentiation frameworks in deep-learning platforms.
To ensure the existence of the gradient and the validity of the computation, the objective function must be continuously differentiable, i.e., $f \in C^1(\mathbb{C}^N, \mathbb{R})$.

\subsection{Objective Function Design}\label{subsec:object}

\subsubsection{Principal Square Root Function $\sqrt{\cdot}$ }\label{subsubsec:sqrt}
For any $z \in \mathbb{C} \setminus (-\infty, 0]$, its square root is defined as:
\begin{equation}
    \sqrt{z}:=e^{\frac12(\ln|z|+j\arg z)}.
\end{equation}
Here, $\arg z$ is the principal value of the argument. Within this domain, $\sqrt{z}$ is holomorphic.

\subsubsection{Amplitude Matching Objective Function $f^p_{\mathrm{main},i}(\boldsymbol{z})$ }
Given the measurements $y_i \in \mathbb{R}$, the corresponding sampling vectors $\boldsymbol{u}_i \in \mathbb{C}^N$, the complex coefficient vector $\boldsymbol{z} \in \mathbb{C}^N$ to be optimized, and the power parameter $p \in \mathbb{R}+$, the amplitude matching objective function is defined as
\begin{equation}
    f^p_{\mathrm{main},i}(\boldsymbol{z}) = \left |y_i - |\boldsymbol{u}^{*}_i \boldsymbol{z}|\right |^p = \left |y_i - \sqrt{\overline{\boldsymbol{z}}^{\top} (\boldsymbol{u}_i\boldsymbol{u}^{*}_i) \boldsymbol{z}} \right |^p,
\end{equation}
where $\overline{\boldsymbol{z}}^{\top} (\boldsymbol{u}_i\boldsymbol{u}^{*}_i) \boldsymbol{z} \in \mathbb{R}{\geq 0}$.
When $\boldsymbol{u}_i \boldsymbol{z} \neq 0$, the quantity under the square root is a positive real number, satisfying the holomorphic domain condition in \cref{subsubsec:sqrt}; when $\boldsymbol{u}_i \boldsymbol{z} = 0$, the objective function degenerates to $\left|y_i\right|^p$, forming a zero-measure set under the assumption of continuous probability.

\subsubsection{Power-Normalization Objective Function $f^p_{\mathrm{norm}}$ }
To ensure that the complex coefficient vector satisfies the power-normalization constraint (\textit{cf.} \cref{eq:reg}), the power normalization objective function is defined as
\begin{equation}
    f^p_{\mathrm{norm}}(\boldsymbol{z}) = |\boldsymbol{z}^* \boldsymbol{z}-1|^p.
\end{equation}

\subsection{Large-Step Optimization Algorithm}\label{subsec:Algorithm}

Nonconvex optimization often get trapped in poor local minima.
Recent studies indicate that a large-step-size strategy helps to escape local minimum basins \cite{10.5555/3495724.3495945}, traverse narrow attraction regions \cite{pmlr-v202-mohtashami23a}, and thus increase the likelihood of approaching a global optimum.
This strategy also induces implicit regularization, guiding the optimizer toward flatter minima \cite{wang2023good}.

The objective function has two characteristics: first, twin‑image ambiguity yields two solutions and produces a more intricate landscape, especially in the presence of noise; second, the power-normalization constraint limits the moduli of the complex coefficients, so optimization near a solution primarily adjusts phases.
Near the optimum, the objective function is relatively flat with respect to phase, while a pseudo-solution distorts the FF pattern and produces a much steeper objective-function landscape in its vicinity.


In nonconvex optimization, an improper step size can cause the process to stall in local optima or to skip over critical points, so an optimizer that intuitively reflects step-size adjustments is required.
The AdaMax optimizer uses $L^\infty$-norm gradient updates, which not only simplify computational complexity but also avoid zero-initialized bias, thereby improving early-iteration stability.
The update magnitude of each parameter is explicitly bounded by the step size $\alpha$, making step-size tuning more intuitive and effective.
In our numerical examples, we choose a large step size $\alpha = 1$ to ensure the optimizer has sufficient “momentum” to escape the local minimum associated with spurious solutions and explore both solutions.
Additionally, to accelerate parameter search, we employ a higher exponent $p = 3$ in the objective, which is defined as:
\begin{equation}
  f_{\mathrm{tot},3} = \sum_{i=1}^{M} f_{\mathrm{main}, i}^3 + \gamma \sum_{i=1}^{M} f_{\mathrm{norm},i}^3.
\end{equation}
This design rapidly amplifies the objective near suboptimal solutions, guiding the optimizer to discard spurious minima more quickly.
In this work, we set the hyperparameter $\gamma = 1$, though this value should be tuned according to the specific task.

\section{Experiments and Metrics}
\subsection{Experimental Setup}
This paper will comprehensively validate the principles, feasibility, robustness, and efficiency of the proposed method through a series of systematic numerical experiments, and this section provides the experimental setup. The overall experimental design follows the following logic:
\begin{enumerate}
    \item \textbf{Loss Landscape}: Visualize the optimization trajectories and loss landscape features via low-dimensional projection to elucidate the method's underlying principles, and subsequently select the optimal optimizer and learning rate.
    
    \item \textbf{Large Number of Modes}: Assess the method's capability to handle large-scale modal analysis problems by increasing the number of electromagnetic modes (via a larger waveguide radius) while maintaining acceptable FF measurement distances.
    
    \item \textbf{Robustness Evaluation}:
    \begin{enumerate}
        \item First, visualize the physical impact of the defined SNR on the AF and FF distributions.
        \item Second, quantitatively evaluate the method's robustness by testing it under various combinations of $\mathrm{SNR}_\mathrm{AF}$ and $\mathrm{SNR}_\mathrm{FF}$.
        \item Finally, investigate the impact of the number of sampling points on noise suppression and reconstruction accuracy.
    \end{enumerate}
    
    \item \textbf{Optimization Method Comparison}: Benchmark the computational cost and convergence accuracy against similar non-convex optimization methods to validate the proposed method's practicality and performance advantages.
\end{enumerate}


All experiments are based on a circular waveguide model operating at a frequency of $f = 2.45~\mathrm{GHz}$, with a corresponding wavelength of $\lambda = c / f$. The radiated power from the waveguide aperture is fixed at $P_{\mathrm{rad}} = 1~\mathrm{W}$. The waveguide radius, $a$, which determines the number of supported electromagnetic modes, is varied in each case. To satisfy the FF condition, the sampling distance $r$ from the port center is set to $r = \lfloor 8a^2 / \lambda \rfloor + 1$, where $\lfloor \cdot \rfloor$ denotes the floor function. The AF points are randomly distributed on the waveguide's aperture plane, while the FF points are randomly distributed on a hemispherical surface.

\subsubsection{Loss Landscape}
To visualize the optimization trajectories, we configured a waveguide with $a = 1.2\lambda \approx 0.147~\mathrm{m}$, supporting $N = 16$ modes, and sampled the FF at $r = 2~\mathrm{m}$. The number of samples was set to $M_{\mathrm{AF}} = M_{\mathrm{FF}} = 2(N - 1)$. We compared the performance of the Adam and AdaMax optimizers with step sizes $\alpha = 0.001, 0.01, 0.1, 0.5, 1$, each experiment for $10^5$ iterations. Based on these results, AdaMax with a learning rate of $\alpha = 0.5$ was used for all subsequent experiments.
\subsubsection{Large Number of Modes}
To assess the algorithm's performance on high-order mode problems, we modeled a waveguide with $a = 3\lambda \approx 0.367~\mathrm{m}$, supporting $N = 93$ electromagnetic modes. The FF was sampled at $r = 9~\mathrm{m}$ with $M_{\mathrm{AF}} = M_{\mathrm{FF}} = 2N(N - 1)$ samples. The optimization was run for  $3\times10^4$ iterations.

\subsubsection{Robustness Evaluation}
To intuitively demonstrate the physical impact of the SNR, we analyze the field distributions at a fixed azimuthal angle of $\varphi = 45^\circ$ in both the AF and FF domains.
In the first case, to isolate the effects of background noise, the AF was kept noiseless while the FF was subjected to noise levels corresponding to $\mathrm{SNR}_\mathrm{FF}$ of 10, 20, 30, 40, 50, and 60 dB, in addition to a noiseless reference. The resulting FF amplitude distribution is plotted against the zenith angle $\theta$ to observe the degradation of the radiation pattern.
In the second case, to isolate the effects of geometry noise, the FF was kept noiseless while the $\mathrm{SNR}_\mathrm{AF}$ was varied across the same levels. The AF amplitude distribution is then plotted against the radial coordinate $\rho$ to demonstrate the distortion of the AF profile. 
Together, this comparative analysis serves to illustrate the physical meaning of our defined SNR and the distinct impact of noise in each observation domain.
To investigate the impact of noise on reconstruction performance, different combinations of $\mathrm{SNR}_{\mathrm{AF}}$ and $\mathrm{SNR}_{\mathrm{FF}}$ are considered at 10, 20, 30, 40, 50, and 60 dB.
The waveguide radius $a$ ranges from $0.4\lambda$ to $1.8\lambda$, corresponding to mode numbers $N = 2,3,6,10,13,16,18,23,24,28,31,34$. The number of sampling points is set as $M_{\mathrm{AF}} = M_{\mathrm{FF}} = 2(N - 1)$.
Each experiment is iterated $3\times10^5$ times and repeated five times, and the average values of $\mathrm{MRE}_{\mathrm{Modulus}}$ and $\mathrm{MAE}_{\mathrm{Phase}}$ are reported.

We further investigated the effect of the number of sampling points on noise suppression. The waveguide radius was set to $a = 1.1\lambda \approx 0.134~\mathrm{m}$, corresponding to  $N=13$ electromagnetic modes. The settings of $\mathrm{SNR_{AF}}$ and $\mathrm{SNR_{FF}}$ were the same as those indicated by the blue marker (Exp.) in \cref{fig:noise_vis}. The number of sampling points was configured as  $M_{\mathrm{AF}} = M_{\mathrm{FF}} = 2(N - 1), 6(N - 1)^2, 6^2(N - 1)^2, 2\times6^2(N - 1)^2, 3\times6^2(N - 1)^2,4\times6^2(N - 1)^2, 5\times6^2(N - 1)^2, 6\times6^2(N - 1)^2$. Each experiment is iterated $3\times10^5$ times and repeated 40 times; the results are presented using box plots.

\subsubsection{Comparison of Optimization Methods}

To compare the performance of different optimization methods in terms of reconstruction accuracy and computational cost, we benchmark our approach against the SPGD algorithm  \cite{Lu:13} and the hybrid optimization framework proposed in \cite{Li:17}. Both studies were conducted on tasks involving six LP modes.
In our experiments, the aperture radius is set to $a = 0.8\lambda \approx 0.0979~\mathrm{m}$, corresponding to $N = 6$ electromagnetic modes, and the observation distance is $r = 1~\mathrm{m}$.
The number of sampling points is set as $M_{\mathrm{AF}} = M_{\mathrm{FF}} = 2(N - 1)$.


\subsection{Performance Metrics}

\subsubsection{$\mathrm{MRE}_{\mathrm{Modulus}}$}
Owing to the power-normalization constraint, the modulus $|z_i|$ for each mode decreases progressively as the number of modes increases. Consequently, we employ the mean relative error (MRE) between the ground truth modulus $|z_i|$ and the retrieved modulus $|\hat{z_i}|$ to evaluate the reconstruction efficacy of the modal power distribution. It is calculated as follows:
\begin{equation}
  \mathrm{MRE}_{\mathrm{Modulus}} = \frac{1}{N}\sum_{i=1}^N \left|\frac{|z_i| - |\hat{z_i}|}{|\hat{z_i}|}\right|.
\end{equation}

\subsubsection{$\mathrm{MAE}_{\mathrm{PhaseDiff}}$}
Since a global phase shift $z_i \mapsto z_i e^{j\theta} $ does not affect relative phases,  we establish a common phase reference by fixing the phase of the fundamental mode to zero, and all other modal phases  are then rotated accordingly to align with this reference. The accuracy of the relative phase distribution is subsequently evaluated by calculating the Mean Absolute Error (MAE) between the ground truth phases $\phi_i$ and retrieved phases $\hat{\phi_i}$. It is calculated as follows:
\begin{equation}
  \mathrm{MAE}_{\mathrm{PhaseDiff}} = \frac{1}{N} \sum_{i=1}^{N} \left|\phi_i - \hat{\phi}_i\right|.
\end{equation}

\subsubsection{$\mathrm{MAE}_{\mathrm{Modulus}}$}
Existing optimization-based methods typically evaluate performance using absolute error. Although these methods neglect the power normalization constraint, which leads to a significant overestimation of power distribution accuracy as the number of modes increases, we additionally introduce the MAE between the true amplitudes $|z_i|$ and the reconstructed amplitudes $|\hat{z_i}|$ to evaluate power recovery results. This metric is used for performance comparison in \cref{sec:perform} and is defined as

\begin{equation}
  \mathrm{MAE}_{\mathrm{Modulus}} = \frac{1}{N}\sum_{i=1}^N ||z_i| - |\hat{z_i}||.
\end{equation}



\section{Results}

\subsection{Loss Landscape}

\begin{figure*}[!t]
    \centering
    \includegraphics[width=1\linewidth]{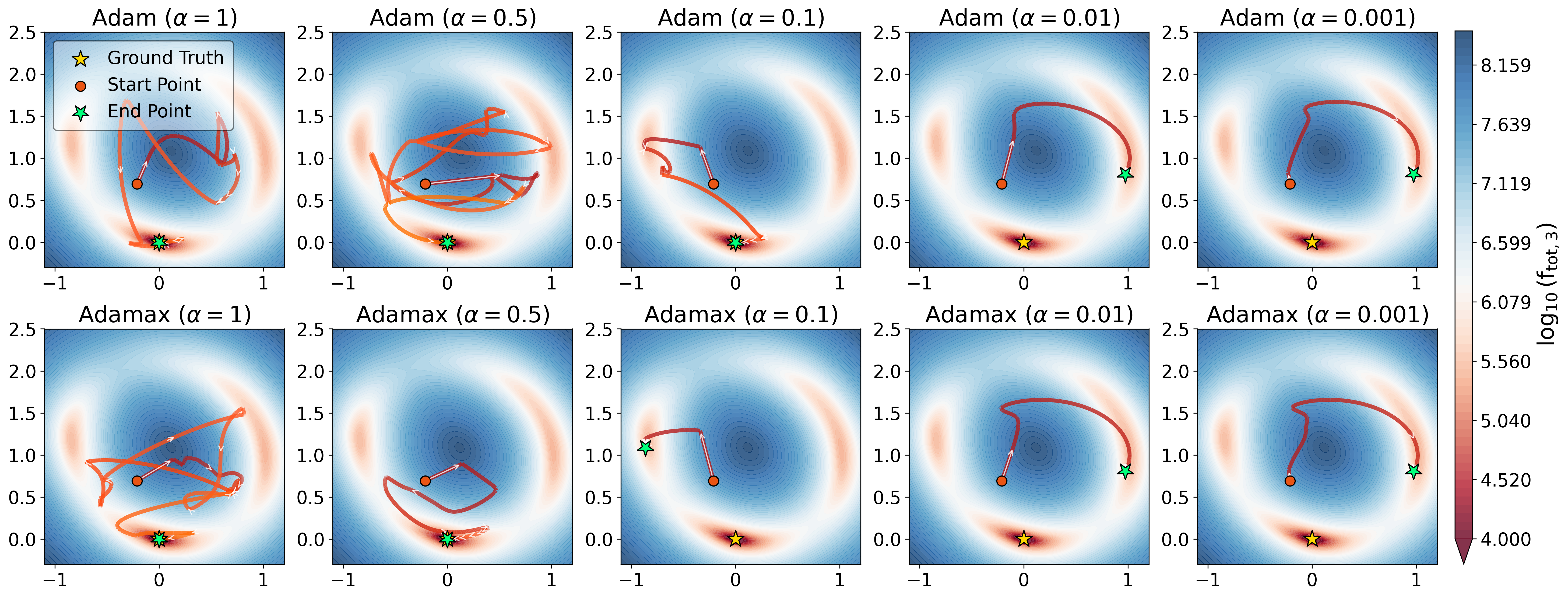}
    \caption{The Gaussian-smoothed projection results of optimization trajectories with different step sizes and optimizers are visualized on the PCA plane, together with the corresponding loss landscape.The trajectories evolve from a shared start point (red dot) toward an endpoint (green star), with the red-to-orange color gradient marking iterative progress towards the ground truth (gold star). The white arrows indicate the direction of optimization. The underlying loss landscape is depicted by the colored isograph. The trajectories with larger step sizes exhibit a tendency to traverse across local minima.}
    \label{fig:landscape}
\end{figure*}

\begin{figure}[!t]
    \centering
    \includegraphics[width=1\linewidth]{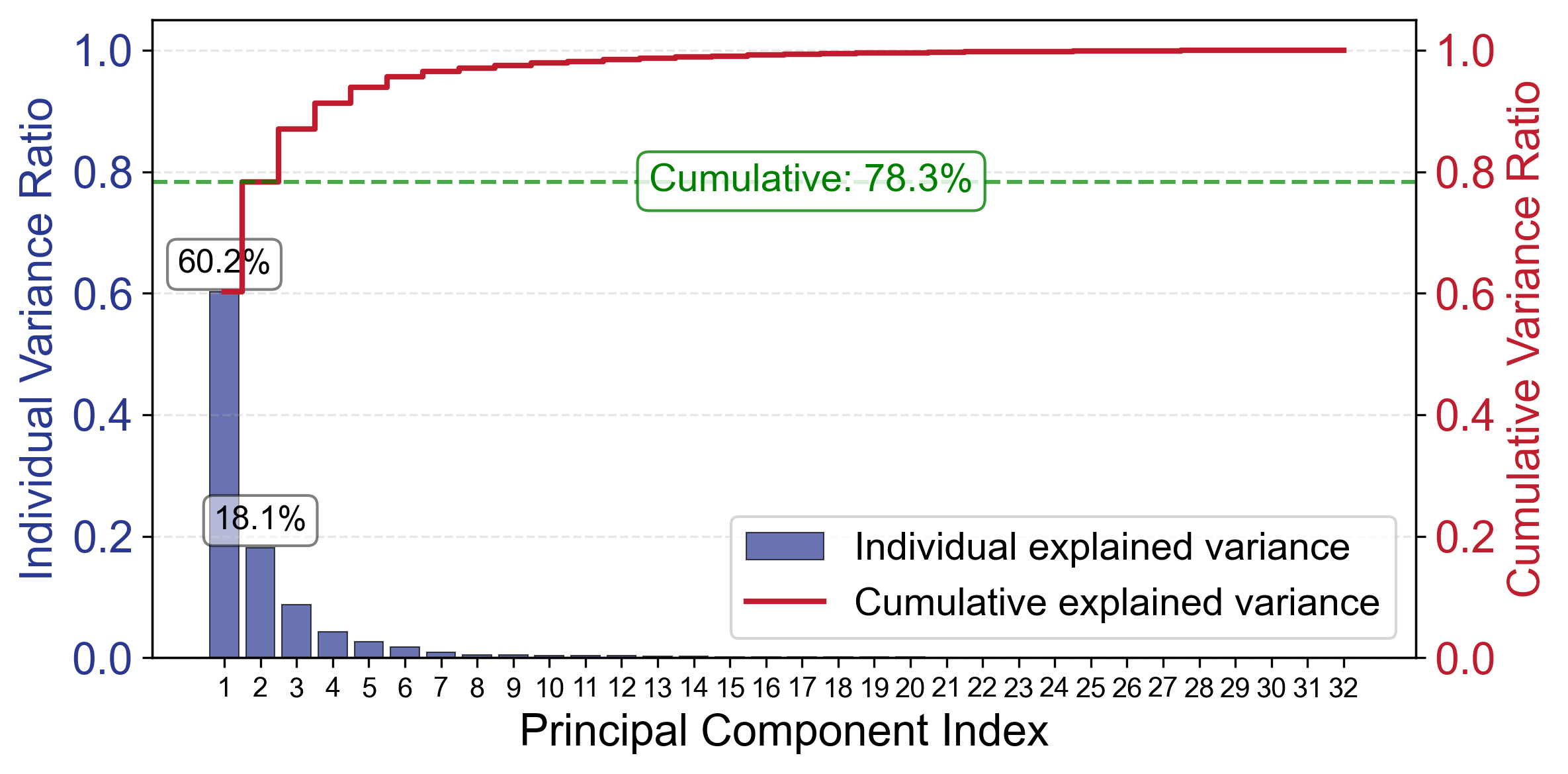}
    \caption{PCA of parameter evolution across $10^5$ iterations. The bar chart shows the individual explained variance for individual principal component, while the red line indicates the cumulative explained variance. The first two principal components account for 78.3\% of the total variance, capturing the dominant directions of parameter evolution.}
    \label{fig:PCA}
\end{figure}

To investigate the role of large-step updates in non-convex optimization, we analyzed trajectories under varying step sizes using adaptive optimizers. 
The trajectories were visualized through principal component analysis (PCA) \cite{FRS1901LIIIOL}, which projects high-dimensional parameter updates onto orthogonal directions of maximal variance for compact representation. Across $10^5$ iterations, parameter updates were projected onto a unified PCA space, as shown in \cref{fig:PCA}, where the first two components captured 78.3\% of the variance, revealing dominant evolution directions.

Small-step trajectories evolved quasi-continuously, whereas large-step trajectories consisted of discrete jumps. Gaussian smoothing suppressed high-frequency fluctuations, highlighting the main optimization directions and global parameter evolution.
\cref{fig:landscape} highlights a fundamental trade-off: while small steps offer rapid convergence, they risk confinement to local minima. Large steps, however, enable robust global exploration by navigating multiple optima to ultimately locate the global solution.
The large-step-size strategy drives the optimizer to broadly traverse the multiple significant basins of attraction within the loss landscape; once the optimization path enters the steep basin defined by a correct solution, its large and directionally clear gradient field dominates the dynamics, leading to a rapid and deterministic convergence.
We recommend the AdaMax optimizer for the rapid and stable task adaptation it provides, a direct benefit of its infinity norm strictly bounding the update step by the step size. Based on the empirical results shown graphically, a step size of $\alpha = 0.5$ is adopted for all subsequent experiments.

\subsection{Large number of modes}\label{subsec:large_number}

\begin{figure}[!tbp]
    \centering
    \includegraphics[width=1\linewidth]{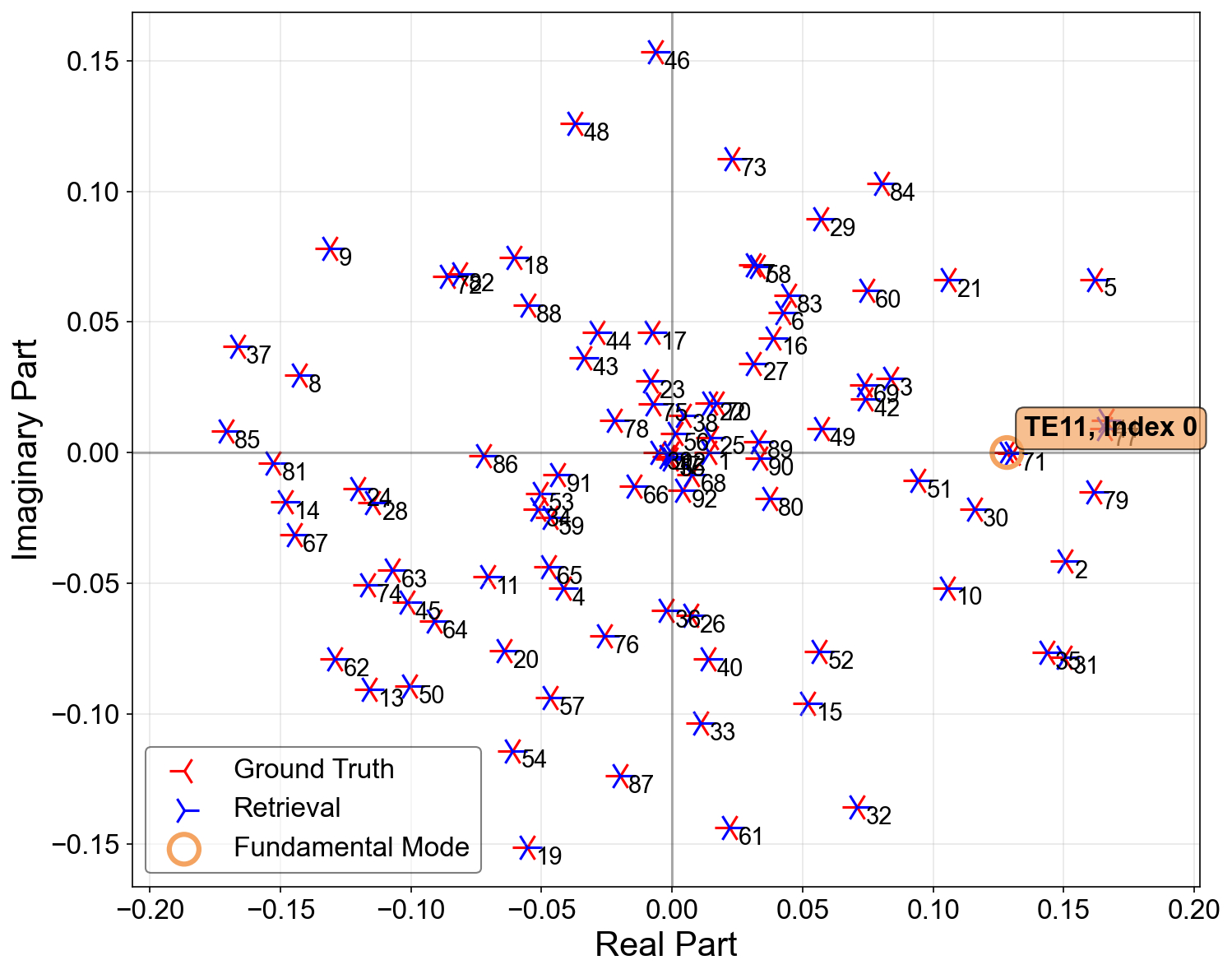}
    \caption{Comparison of retrieved and ground truth modal complex coefficients for 93 modes. The retrieved complex coefficients (blue scatters) and their ground truth values (red scatters) are plotted on the complex plane. All coefficients are referenced to the fundamental TE11 mode, whose phase is fixed at zero (orange circle). The position of each point is determined by its normalized modulus and relative phase. }
    \label{fig:large_result}
\end{figure}

\begin{figure*}[!t]
    \centering
    \begin{minipage}[c]{0.48\linewidth}  
        \centering
        \includegraphics[width=\linewidth]{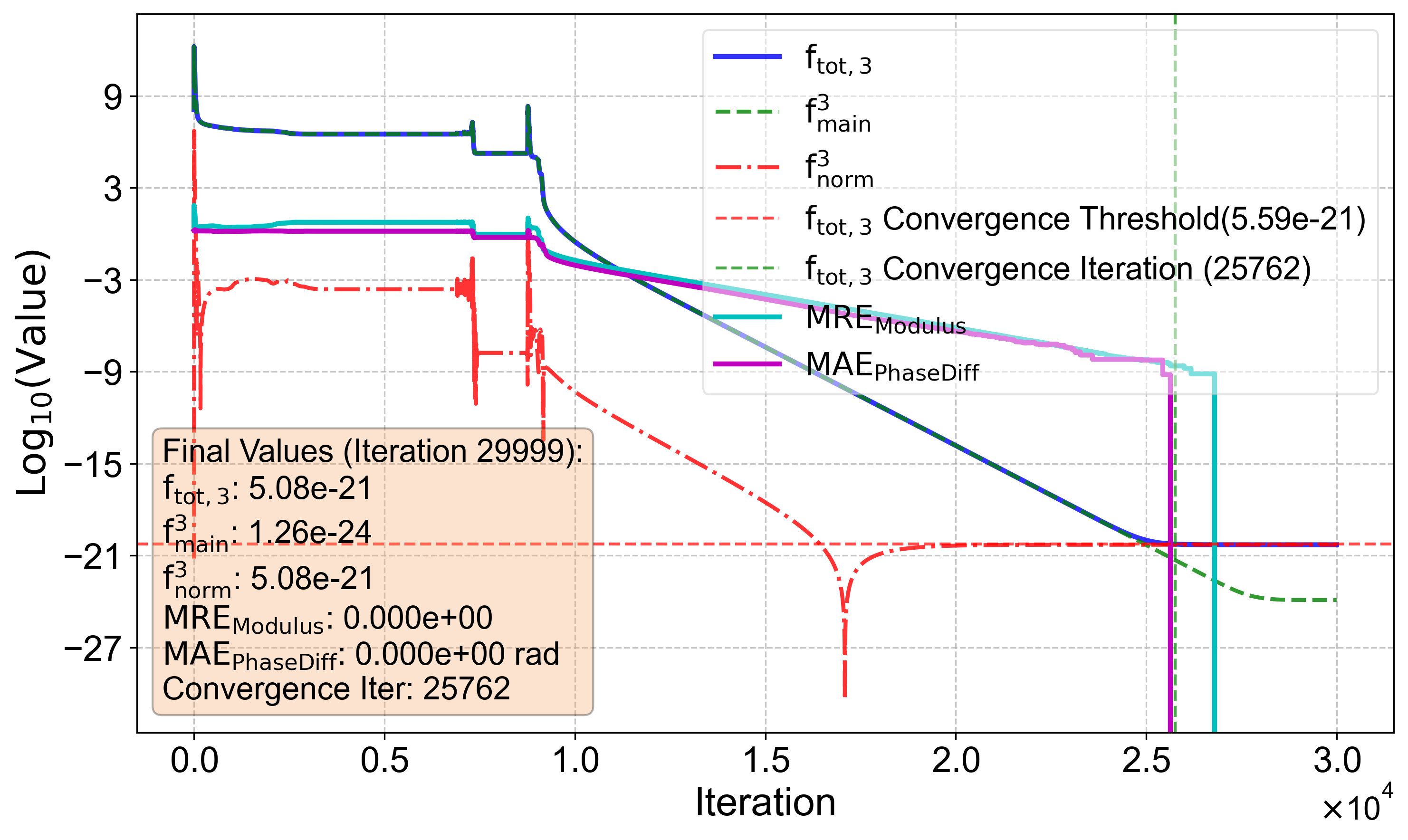}
        \caption{Convergence dynamics of various key metrics evolving with the number of iterations for mode analysis with a large number of modes. The plot displays the evolution of the total loss $f_{\mathrm{tot},3}$, its components $f_{\mathrm{main}}^3$ and $f_{\mathrm{norm}}^3$ ,  and the evaluation metrics $\mathrm{MRE_{Modulus}}$ and $\mathrm{MAE_{PhaseDiff}}$. The convergence iteration and threshold for $f_{\mathrm{tot},3}$ are annotated, along with the final values at the last iteration.}
        \label{fig:large_analysis}
    \end{minipage}
    \hfill
    \begin{minipage}[c]{0.48\linewidth}  
        \centering
        \includegraphics[width=0.75\linewidth]{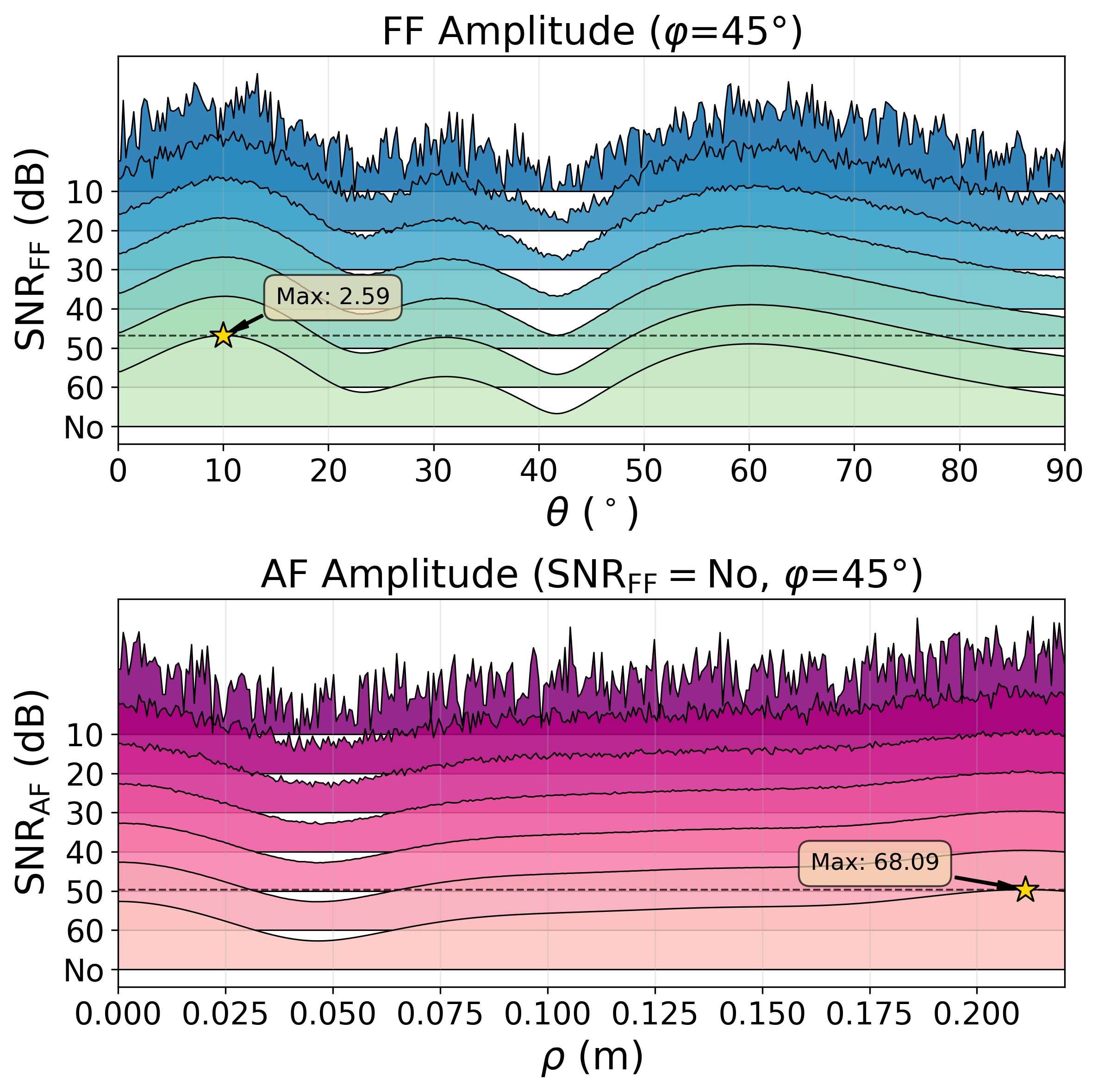}
        \caption{Comparative analysis of noise impact on FF and AF distributions.  Ridgeline plots showing the degradation of the FF radiation pattern (top) and AF amplitude (bottom) with decreasing SNR. The analysis is performed at a fixed azimuthal slice ($\varphi = 45^\circ$), with markers indicating the peak amplitude of the noiseless reference case.}
        \label{fig:noise_ridge}
    \end{minipage}
\end{figure*}

\begin{figure*}[!t]
    \centering
    \includegraphics[width=\linewidth]{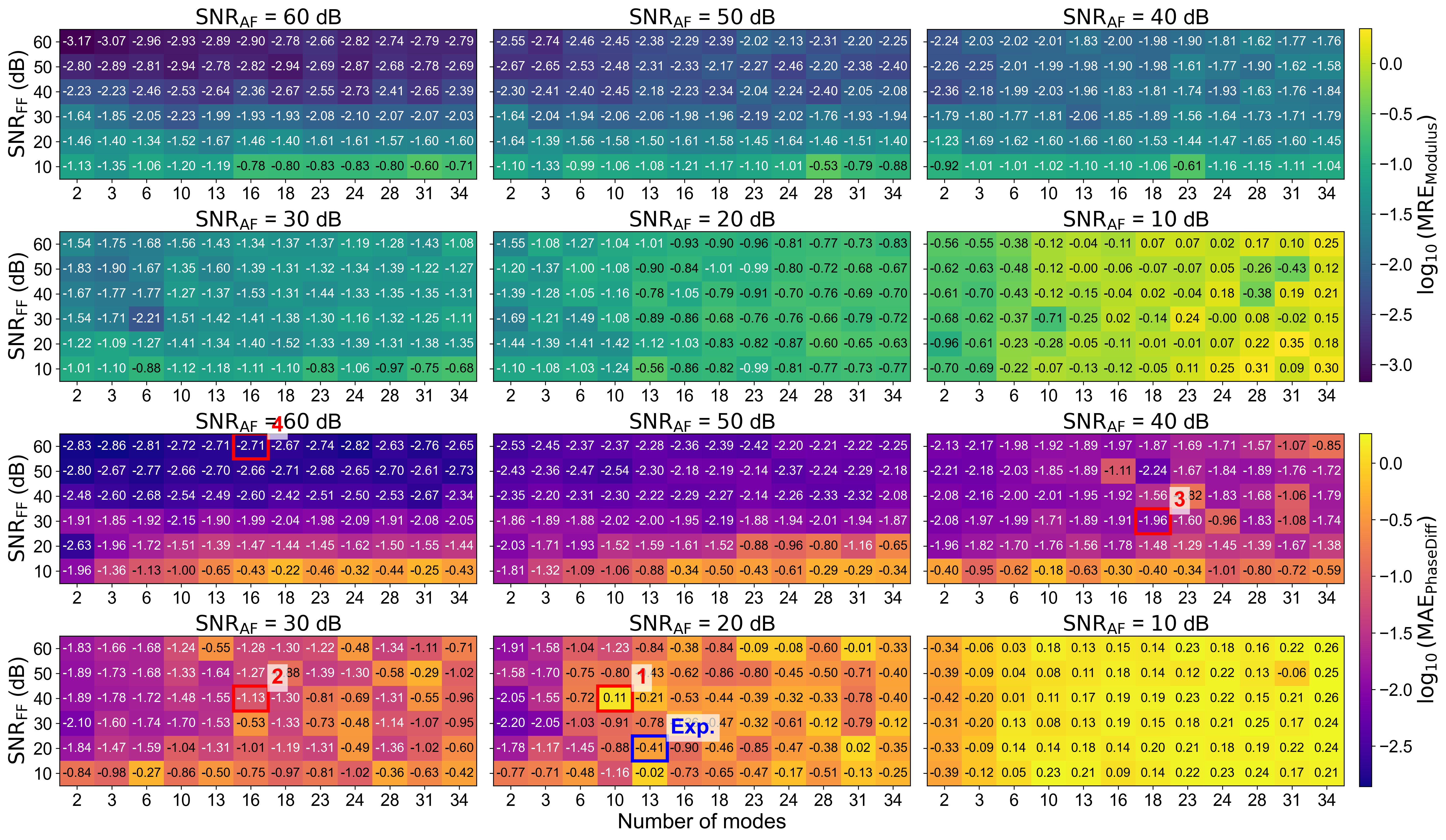}
    \caption{Robustness evaluation of the modal analysis. This heatmap matrix shows the variation of $\mathrm{MRE_{Modulus}}$ and $\mathrm{MAE_{PhaseDiff}}$ with $\mathrm{SNR}_\mathrm{AF}$ and $\mathrm{SNR}_\mathrm{FF}$ for different numbers of modes. Darker colors represent smaller errors and correspondingly higher accuracy in the mode analysis. As the $\mathrm{MAE_{PhaseDiff}}$ is a more representative metric of the analysis accuracy, the corresponding heatmaps are annotated to highlight key parameter conditions.  The red numbers (1-4) mark four cases with different error levels on the $\log_{10}$ scale for typical result analysis, and the blue marker (Exp.) indicates the conditions for a subsequent study on the effect of the number of sampling points.}
    \label{fig:noise_vis}
\end{figure*}

Modal analysis for waveguides supporting a large number of modes presents a significant challenge. This section demonstrates our method's efficacy on such a large-scale problem, analyzing 93 modes at the designated operating frequency and a conventional FF measurement distance. The results, shown in \cref{fig:large_result}, indicate that the method retrieves the modal complex coefficients with high precision, achieving near-perfect agreement with the ground truth values.

The convergence dynamics, including the loss function and custom evaluation metrics, are visualized in \cref{fig:large_analysis}. The plot shows that during an initial exploratory phase of approximately 10,000 iterations, the metrics do not significantly decrease. This behavior is consistent with the global search pattern identified in \cref{fig:landscape} for optimizers using large learning rates (e.g., $\alpha=0.5$ or $1.0$). Following this phase, the metrics begin a steady decline, indicating convergence to the global optimum. This two-phase dynamic is highly consistent with the compressed, smooth trajectory features observed in the loss landscape analysis. The optimization achieves machine precision for the $\mathrm{MRE_{Modulus}}$ at iteration 25,762, with the $\mathrm{MAE_{PhaseDiff}}$ reaching the same precision shortly thereafter. This result validates the method's numerical accuracy and reliable convergence for large-scale modal analysis problems.


\begin{figure*}[!tbp]
    \centering
    \includegraphics[width=1\linewidth]{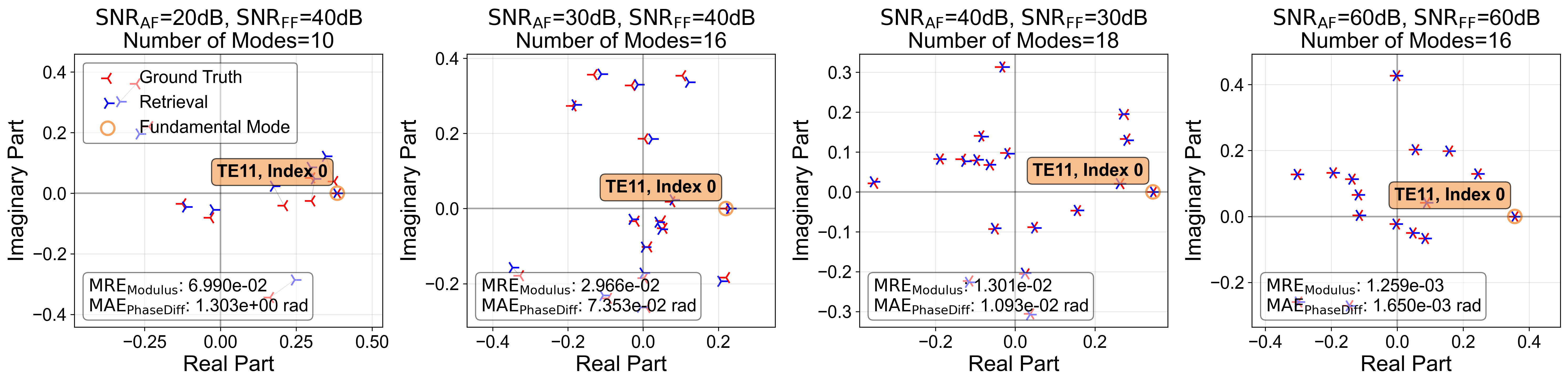}
    \caption{Comparison of Retrieved and Ground Truth Modal Complex Coefficients for four typical results.  The retrieved complex coefficients (blue scatters) and their ground truth values (red scatters) are plotted on the complex plane. The panels show the median result from multiple trials, with corresponding error metrics $\mathrm{MRE_{Modulus}}$ and $\mathrm{MAE_{PhaseDiff}}$ annotated for each case.}
    \label{fig:noise_case}
\end{figure*}

\begin{figure}[!tbp]
    \centering
    \includegraphics[width=0.8\linewidth]{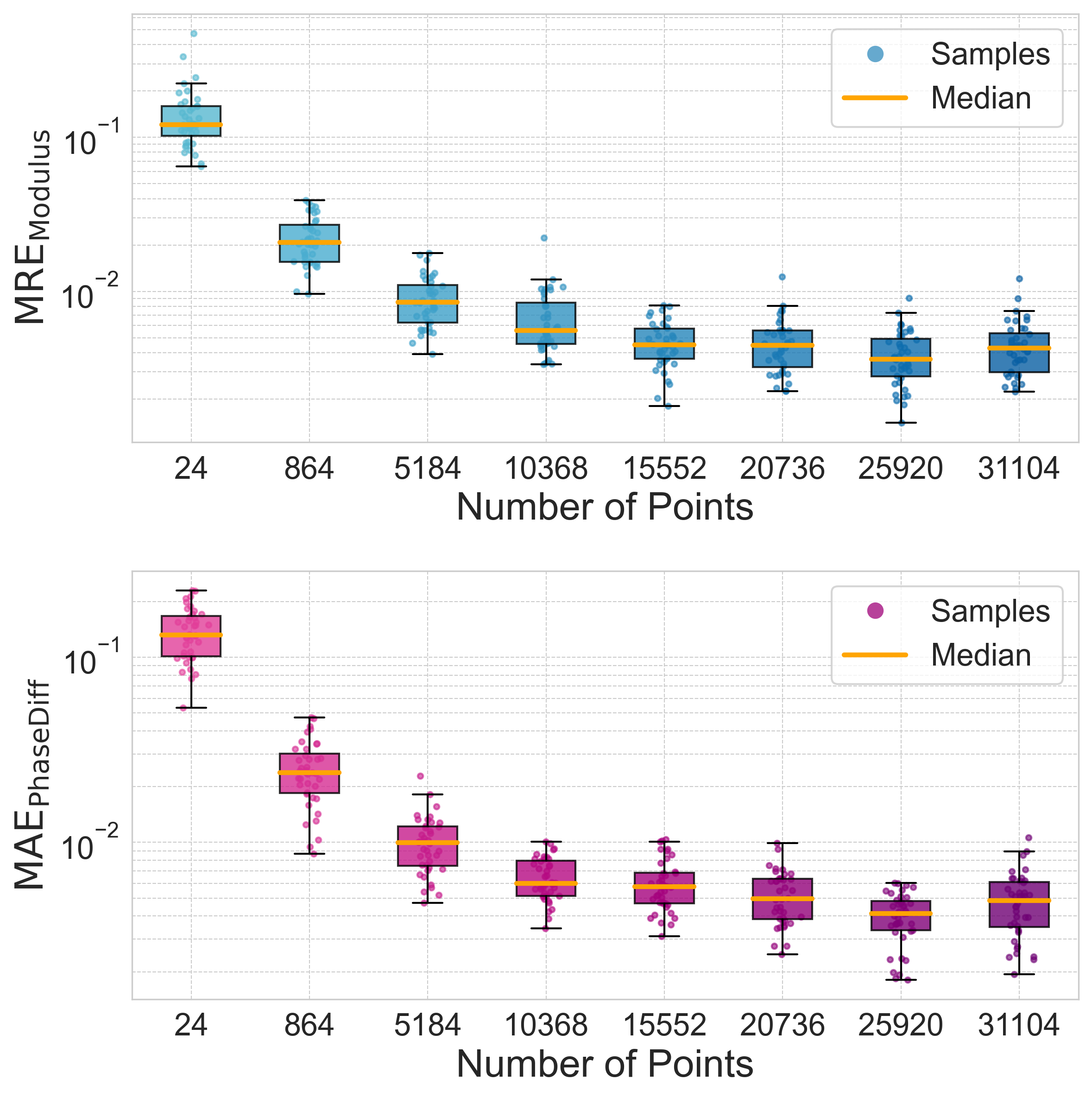}
    \caption{Relationship between the number of sampling points and mode analysis accuracy. This figure uses box plots to show the effect of the number of sampling points on $\mathrm{MRE_{Modulus}}$ and $\mathrm{MAE_{PhaseDiff}}$. In each box plot, the colored dots are the raw data points, the box spans from the first quartile ($\mathrm{Q_1}$) to the third quartile ($\mathrm{Q_3}$), the orange line inside is the median ($\mathrm{Q_2}$), and the whiskers mark the data range, defined as 1.5 times the interquartile range ($\mathrm{IQR = Q_3 - Q_1}$) from the box edges.}
    \label{fig:noise_err}
\end{figure}

\subsection{Robustness Evaluation}

\afterpage{
\begin{figure*}[!tbp]
    \centering
    \includegraphics[width=1\linewidth]{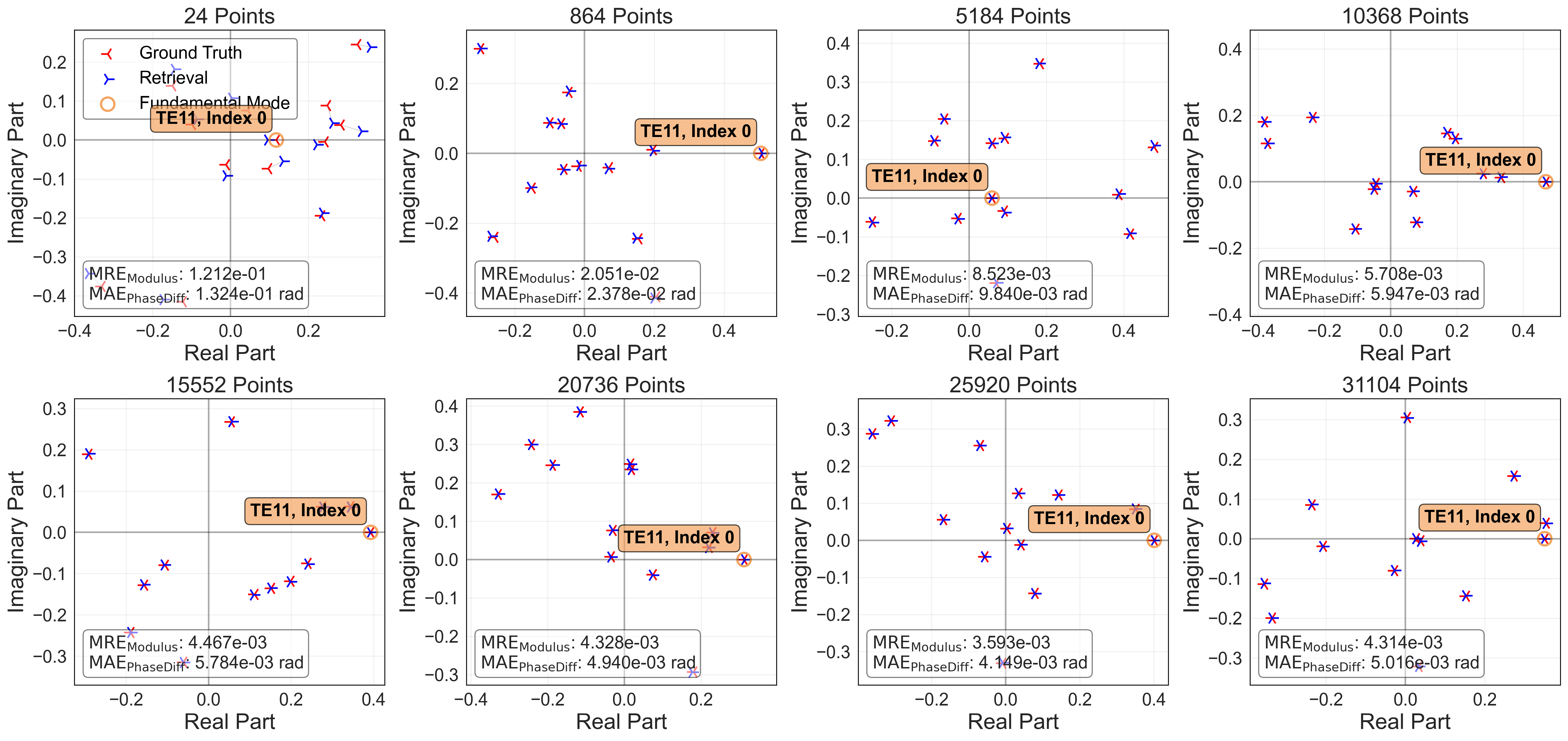}
    \caption{Mode analysis results for the median $\mathrm{MAE_{PhaseDiff}}$ at different numbers of sampling points.  The retrieved complex coefficients (blue scatters) and their ground truth values (red scatters) are plotted on the complex plane.  Each plot is annotated with the $\mathrm{MRE_{Modulus}}$ and $\mathrm{MAE_{PhaseDiff}}$ for the corresponding median result.}
    \label{fig:noise_exp}
\end{figure*}
}

This section aims to systematically evaluate the robustness of the proposed method. We first test the algorithm's performance across a broad parameter space of SNRs, and then further verify the effectiveness of increasing the number of sampling points to mitigate the impact of noise and enhance analysis accuracy.

We have mathematically defined the $\mathrm{SNR}_\mathrm{AF}$ and $\mathrm{SNR}_\mathrm{FF}$ in \cref{subsec: noise}. To intuitively demonstrate the physical meaning of these definitions, we visualize the degradation of the FF radiation pattern and AF amplitude distribution under various SNR levels, as shown in \cref{fig:noise_ridge}. It is worth emphasizing that the SNR range tested in this study (e.g., 10 dB or 20 dB) represents extreme noise conditions not typically encountered in a standard laboratory environment. This setup is designed to rigorously test the performance limits of our method under adverse conditions. Nevertheless, an $\mathrm{SNR}_\mathrm{AF}$ of approximately 20 dB can still be caused by factors such as imperfections in the electromagnetic geometric structure, thus giving this analysis practical relevance.

To systematically evaluate the algorithm's robustness, we conducted exhaustive experiments to study the effect of different SNR combinations on the mode analysis accuracy, with the results shown in \cref{fig:noise_vis}. The results reveal several key trends. First, both error metrics ($\mathrm{MRE_{Modulus}}$ and $\mathrm{MAE_{PhaseDiff}}$) follow a common pattern: the error increases as the number of modes increases or as either $\mathrm{SNR}_\mathrm{AF}$ or $\mathrm{SNR}_\mathrm{FF}$ decreases. Second, we observe a critical performance threshold: when $\mathrm{SNR}_\mathrm{AF}$ is less than or equal to 10 dB, the $\mathrm{MRE_{Modulus}}$ is on the order of $10^0$, indicating that the modulus recovery is unacceptable. However, when $\mathrm{SNR}_\mathrm{AF}$ is greater than 10 dB, the $\mathrm{MRE_{Modulus}}$ drops rapidly below the order of $10^0$, and the results become reliable. Similarly, when $\mathrm{SNR}_\mathrm{FF}$ is below 10 dB, the phase is almost entirely unrecoverable. Finally, a comparison of the two metrics reveals that, under the same SNR and mode count conditions, phase recovery is significantly more difficult than modulus recovery, meaning $\mathrm{MAE_{PhaseDiff}}$ is more sensitive to noise.
Based on the finding that phase recovery is more challenging, we selected four representative error levels from the $\mathrm{MAE_{PhaseDiff}}$ heatmaps for an in-depth case study, as shown in \cref{fig:noise_case}. This analysis provides an intuitive physical interpretation for different orders of magnitude of $\mathrm{MAE_{PhaseDiff}}$: when $\mathrm{MAE_{PhaseDiff}}$ is on the order of $10^0$, the phase recovery for some modes fails; at the order of $10^{-1}$, the phase is mostly recovered but with minor errors; and when the error decreases to the order of $10^{-2}$ to $10^{-3}$, an accurate phase recovery is achieved. For the issue of errors potentially reaching the order of $10^0$ in high-noise environments, a solution can be effectively found by increasing the number of sampling points.

To verify that increasing the number of sampling points can overcome the effects of strong noise, we conducted a test under a challenging condition: 13 modes with both SNRs at 20 dB (see the blue marker in \cref{fig:noise_vis}). As shown in \cref{fig:noise_err}, the experimental results clearly confirm the negative correlation between error and the number of sampling points. Further analysis of the median $\mathrm{MAE_{PhaseDiff}}$ results (\cref{fig:noise_exp}) shows that only 864 sampling points are needed to reduce the phase error to an accurate level on the order of $10^{-2}$, and increasing the samples further continues to improve performance. This body of evidence demonstrates that increasing the sampling density is an effective strategy for ensuring high-accuracy mode analysis results in noisy environments, which further enhances the practicality and scalability of the proposed method.

\subsection{Comparison of Optimization Methods}\label{sec:perform}

\begin{table}[!htbp]
    \caption{Conversion of Evaluation Metrics}
    \label{tab:formula_transform}
    \centering
    \begin{tabular}{ccc}
    \toprule
    \textbf{Method} & $\boldsymbol{\mathrm{MAE}_{\mathrm{Modulus}}}$ & $\boldsymbol{\mathrm{MAE}_{\mathrm{PhaseDiff}}}$\\
    \midrule
    SPGD   & $\mathrm{mean}\!\left(\dfrac{R_a}{2\sqrt{\mathrm{Power\ Contents}}}\right)$
       & $\mathrm{mean}\!\left(\bigl|P_a\bigr|\right)$ \\
    Hybrid & $\mathrm{mean}\!\left(\dfrac{\Delta \overline{\rho^2}}{2\sqrt{\overline{\rho^2}}}\right)$
       & $\mathrm{mean}\!\left(\Delta\overline{\theta}\right)$ \\
    \bottomrule
    \end{tabular}
    \end{table}

\begin{table}[!htbp]
    \caption{Comparison of Accuracy and Time Cost Across Different Methods}
    \label{tab:performance_comparision}
    \centering
    \begin{tabular}{ccc}
    \toprule
    \textbf{Method} & $\boldsymbol{\mathrm{MAE}_{\mathrm{Modulus}}}$ & $\boldsymbol{\mathrm{MAE}_{\mathrm{PhaseDiff}}}$  \\
    \midrule
    SPGD & $ 9.924\times 10 ^{-5} $ & $1.230\times 10^{-4}$  \\
    Hybrid & $ 2.617 \times 10^{-3} $  & $ 8.418\times 10 ^{-2}$  \\
    Ours & $\mathbf{1.633\times 10^{-8}}$& $\mathbf{0}$\\
    \bottomrule
    \end{tabular}
    \end{table}

We compare our proposed method with two related approaches, SPGD \cite{Lu:13} and the hybrid framework \cite{Li:17}, both of which apply optimization strategies for modal analysis. The prior studies analyze six LP modes, while our method addresses six circular-waveguide modes, making the comparison valid.
It should be noted that we do not attempt to reproduce their experimental results.
The data from Table I of \cite{Lu:13} (using only the first two sets; the third corresponds to the twin solution) and Table V of \cite{Li:17} are converted into the metrics $\mathrm{MAE}_{\mathrm{Modulus}}$ and $\mathrm{MAE}_{\mathrm{PhaseDiff}}$ using the formulas in \cref{tab:formula_transform}, with symbols consistent with the original sources.
The comparison, shown in \cref{tab:performance_comparision}, reveals that even without analyzing LP modes, our method achieves a clear accuracy advantage.


In Table II of the SPGD method \cite{Lu:13}, the error for ten LP modes reaches 0.56\%. In \cref{subsec:large_number}, we demonstrate that our approach remains effective when the mode count is large.
The marked success in large-number-of-modes tasks is primarily due to our use of a large-step optimization strategy.
When the number of modes is small, the objective landscape is relatively simple; small steps converge more quickly but are more prone to becoming trapped in local minima.

In terms of computational efficiency, Table II in \cite{Lu:13} reports that the SPGD method requires an average of approximately 2.8 s to successfully optimize a task involving six modes on a standard personal computer (model not specified), though it suffers from twin-image ambiguity. The hybrid optimization method \cite{Lu:13} , on the other hand, takes about 150 s in total on an Intel Core i5-4460 processor, with roughly 140 s spent on the GA stage and 10 s on the SPGD stage.

Based on the Whetstone benchmark for floating-point performance (Apple M2: 42.50 GFLOPs vs. Intel Core i5-4460: 16.28 GFLOPs), we estimated the equivalent runtime across platforms. The proposed method takes approximately 3.0 s on the Apple M2 platform, which corresponds to about 7.8 s on the Intel Core i5-4460. Considering both computational stability and reconstruction accuracy, the proposed method demonstrates a significant advantage in computational cost.



\section{Conclusion}
This paper proposes a large-step nonconvex optimization strategy based on AF and FF amplitudes for modal analysis of multimode waveguides. Theoretical analysis reveals that when using radiation field sampling vectors for mode reconstruction, a twin-image ambiguity problem arises. Furthermore, we demonstrate the consistency between the near-to-far-field transformation and the open-aperture waveguide transformation. By jointly utilizing the AF data, the proposed method performs nonconvex optimization with a large-step updating strategy. Compared with traditional approaches, the proposed method significantly simplifies the measurement setup and reduces system cost. Experimental validation on circular waveguides confirms the effectiveness of the proposed approach.

In both theoretical and experimental studies, the origin and solution of the twin-image ambiguity are systematically analyzed, and the underlying mechanism of the proposed method is elucidated through loss landscape characterization. Additional experiments show that the method remains effective for multimode waveguides with a large number of modes. Noise analysis further demonstrates its strong robustness, and comparative results highlight its superior accuracy and computational efficiency over existing schemes.

In summary, the proposed large-step nonconvex optimization strategy enables accurate reconstruction of both power and relative phase distributions in multimode waveguides, exhibiting excellent performance in terms of accuracy, robustness, and computational efficiency.



Methodologically, this work points to a promising direction for a broad range of optimization problems within computational electromagnetics. The optimization paradigm we have demonstrated—applying a large-step-size strategy to solve deterministic nonconvex inverse problems—has value far beyond modal analysis. Building on the successful navigation of a deterministic loss landscape in this study, our approach offers a systematic blueprint for tackling such challenges: modeling a complex physical problem as a nonconvex optimization task and solving it efficiently with a large-step-size strategy. This pathway is poised to open new and powerful avenues for other challenging tasks in the electromagnetic domain, such as material parameter retrieval, antenna structure optimization, metasurface design, and inverse scattering.

\bibliographystyle{IEEEtran}
\bibliography{IEEEabrv,main}

\newpage

 





\end{document}